\documentclass[%
 reprint,
superscriptaddress,
 aps,
 prx,
]{revtex4-2}

\usepackage[braket, qm]{qcircuit}
\usepackage{braket}
\usepackage{float}

\usepackage{dcolumn}% Align table columns on decimal point
\usepackage{bm}% bold math
\usepackage{mathtools} % for equation alignment
\usepackage{color}  % for coloring edit command
\usepackage[T1]{fontenc}
 \usepackage[capitalize]{cleveref}
 
\definecolor{mscolor}{rgb}{0,0.5,0.5}

\definecolor{tgcolor}{rgb}{0.5,0,0.5}
\definecolor{cpcolor}{rgb}{0.4,0,0.8}
\definecolor{rkccolor}{rgb}{0.5,0.5,0}

\definecolor{lpcolor}{rgb}{0.3,0.3,0.3}

\usepackage{xcolor} 
\newcommand {\rsub}[1]{\textcolor{black}{#1}}

{}
\definecolor{phcolor}{rgb}{0.5,0,0.5}
\newcommand{\UWMPhysics}{Department of Physics, University of Wisconsin-Madison, 1150 University Avenue, Madison, WI, 53706}

\newcommand{\IBM}{Present address: IBM Quantum, T. J. Watson Research Center, 1101 Kitchawan Road, Yorktown Heights, New York 10598}

\newcommand{\UNSW}{
University of New South Wales, Sydney, NSW 2052, Australia}

\newcommand{\Infleqtion}{Infleqtion, Inc., Madison, WI, 53703}

\begin{document}

\title{Variational simulation of the Lipkin-Meshkov-Glick model on a neutral atom quantum computer}

\author{R. Chinnarasu}
\affiliation{\UWMPhysics} 

\author{C. Poole}
\affiliation{\UWMPhysics}

\author{L. Phuttitarn}
\affiliation{\UWMPhysics}

\author{A. Noori}
\affiliation{\UWMPhysics} 
\affiliation{\IBM}

\author{T. M. Graham}
\affiliation{\UWMPhysics} 

\author{S. N. Coppersmith}
\affiliation{\UNSW} \affiliation{\UWMPhysics} 

\author{A. B. Balantekin}
\affiliation{\UWMPhysics} 

\author{M. Saffman}
\affiliation{\UWMPhysics}
\affiliation{\Infleqtion}

\date{\today}% It is always \today, today,

\begin{abstract}
We simulate  the Lipkin-Meshkov-Glick (LMG) model using the Variational-Quantum-Eigensolver (VQE) algorithm on a neutral atom quantum computer. We test the ground-state energy of spin systems with up to 15 spins. Two different encoding schemes are used: an individual spin encoding where each spin is represented by one qubit, and an efficient Gray code encoding scheme which only requires a number of qubits that scales with the logarithm of the number of spins. This more efficient encoding, together with zero noise extrapolation techniques, is shown to improve the fidelity of the simulated energies with respect to exact solutions.    
\end{abstract}

\maketitle

% \tableofcontents

 \section{Introduction}

Scaling up of digital quantum processors for running large circuits requires establishing connectivity between distant qubits. Universal paradigms for enabling arbitrary connectivity rely on either moving quantum states across the processor with a sequence of swap gates\cite{DiVincenzo1995b} or using teleportation\cite{Gottesman1999}. Both of these approaches incur additional resource overhead as long swap chains increase circuit depth, and efficient teleportation depends on the availability of high-fidelity resource states. Physical motion of qubits is a third alternative that was originally introduced for the trapped ion platform\cite{Cirac2000, Kielpinski2002,Moses2023}, and has been successfully implemented with neutral atom qubits for logical processing\cite{Bluvstein2022,Bluvstein2024,Reichardt2024}, as well as in atomic tweezer clocks\cite{Finkelstein2024}. In this work, we \rsub{demonstrate several innovations for quantum circuit operation with neutral atom qubits.} Using both  atom transport for dynamical modification of the connectivity map and rapid optical beam scanning for site-specific optical control we simulate the Lipkin-Meshkov-Glick(LMG) model using the Variational-Quantum-Eigensolver(VQE) algorithm. We show that application of efficient encoding strategies together with noise mitigation techniques\cite{Kandala2019} enables finding ground-state energies with an accuracy beyond that supported by the intrinsic physical gate fidelities.  

The LMG model\cite{Lipkin1965} was introduced in the 1960s as an exactly solvable test bed to benchmark classical computer simulations of many-body physics. 
There have been several recent quantum  simulations of the LMG model. A quantum algorithm to find the ground-state energies of a two-particle system was implemented on the IBM Quantum Experience in \cite{Cervia2021}. The excited states of the LMG model were studied in \cite{Hlatshwayo2022}, where the energy spectrum of up to four particles was obtained. Several versions of hybrid quantum-classical approaches to the LMG model are available \cite{Chikaoka2022,ROmero2022,Hengstenberg2023, Robin2023,Beaujeault-Taudiere2024} as well as extensions to the LMG model by inclusion of a pairing term in the Hamiltonian\cite{Illa2023}. 
Due to the exactly solvable nature of the LMG model Hamiltonian, it can be used to verify eigenenergies predicted on NISQ era quantum processors. In this work, we demonstrate simulation of up to 15 spin versions of the LMG Hamiltonian on a neutral atom quantum computer. Ground-state energies found with the VQE algorithm are compared with exact solutions of the LMG Hamiltonian.

The remainder of the paper is organized as follows. In Sec. \ref{sec.VQE} we present the individual spin and Gray code encoding schemes used for VQE simulation of ground-state energies. 
In Sec. \ref{sec.experiment} the experimental approach is described.  Results from both encoding schemes are presented in Sec. \ref{sec.results}. The results are summarized in Sec. \ref{sec.summary} and additional technical details are provided in Appendices.

\section{Model Hamiltonian and spin encoding}
\label{sec.VQE}

The LMG model is a nuclear shell model consisting of $N$ fermions distributed among two $N$-fold degenerate levels ~\cite{Lipkin1965}. In terms of collective quasi-spin operators $J_0, J_+, J_-$, the model Hamiltonian is given by~\cite{Lipkin1965,Cervia2021}  
\begin{equation}
\label{eq:LMGSU2}
{\tilde H} = \epsilon J_0 + \frac{1}{2} {\tilde V} \left( J_+^2 + J_-^2 \right),
\end{equation}
where $\epsilon$ is the energy level separation and $\tilde V$ is the strength of a tunable interaction which flips pairs of particles between the two energy levels.  In the rest of this paper, we work with the dimensionless Hamiltonian $H\equiv {\tilde H}/\epsilon$ with $V\equiv {\tilde V}/\epsilon$. The LMG model is nontrivial but also exactly solvable, making it attractive for use in testing approximation methods in many-body physics. 
The VQE algorithm\cite{Peruzzo2014} leverages the variational principle to estimate the ground-state energy of a Hamiltonian. Compared to Quantum Phase Estimation which in general requires deep circuits, the VQE algorithm is potentially advantageous for implementation on NISQ processors without full error correction.

\begin{figure*}[!t]
    \centering
\includegraphics[width=1.7\columnwidth]{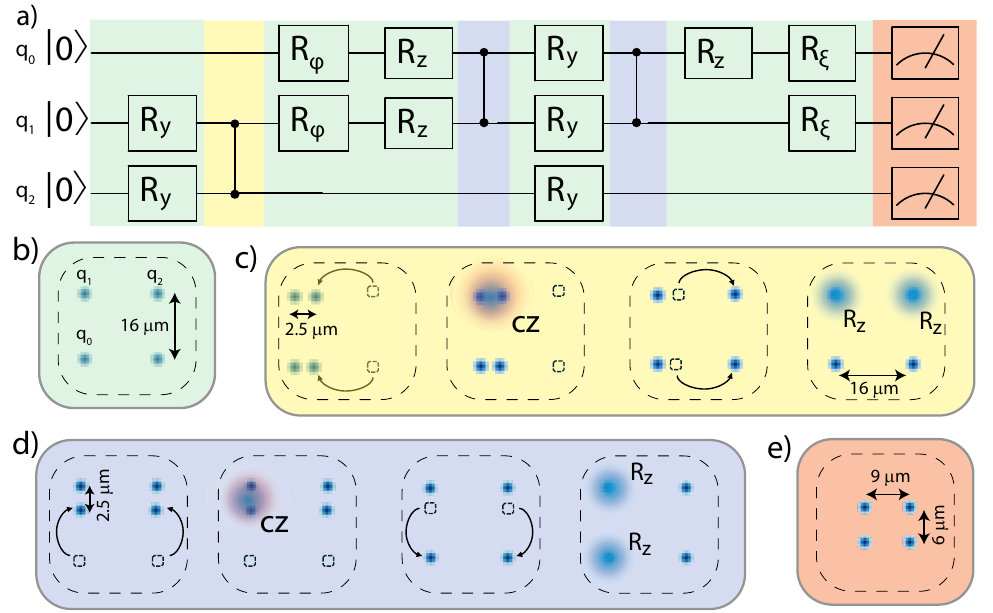}
    \caption{A pictorial  schematic of the three-qubit VQE circuit implementation using optical beam scanning and atom transport. The trap configurations for the different circuit operations are shown in green, yellow, blue, and orange frames. 
    a) Shows a circuit diagram for three qubit VQE state preparation using Gray code encoding. 
    b) An atom spacing of $16~\mu\rm m$  is used for initialization and single qubit operations.
    c) The atoms are moved to a horizontal spacing of $2.5~\mu\rm m$, and Rydberg light is focused on the atom pair in the top row to implement a $\sf CZ$ gate between $q_1, q_2$.  The atoms are then moved apart for local ${\sf R}_z$ correction pulses. 
    d) The atoms are moved to a vertical spacing of $2.5~\mu\rm m$, and Rydberg light is focused on the atom pair in the left column to implement a $\sf CZ$ gate between $q_0, q_1$. The atoms are then moved apart for  local ${\sf R}_z$ correction pulses. 
    e) The atoms are placed on a grid with $6\times9~\mu\rm m$ spacing for readout. This rectangular spacing was chosen to minimize heating from beating between AOD frequencies due to trap crosstalk from scattering and reflections in the trap beam path. 
}
    \label{fig:Schematic}
\end{figure*}

To implement the VQE algorithm, an ansatz wave function $\ket{\psi}$ is selected as a function of a number of variational parameters. The expectation value of the Hamiltonian $\bra{\psi} H \ket{\psi}$ is evaluated and the variational parameters are adjusted to minimize the observed energy.  To implement the algorithm on a quantum computer, the Hamiltonian and trial wave function must be mapped onto the computational basis of the quantum register. To evaluate the expectation value of the Hamiltonian, we decompose it as $H = \sum_{i} a_i P_i$, where $P_i$ are the products of Pauli operators and $a_i = {\rm Tr}(HP_i^{\dagger})/{\rm dim}(H)$. By linearity of expectation values, $\bra{\psi} H \ket{\psi} = \sum_{i} a_i \bra{\psi} P_i \ket{\psi}$. For the results presented in this paper we have set $V=1$ when evaluating the weights $a_i$. We can measure a Pauli operator by measuring each qubit in the basis indicated by the operator. For example, to measure $\sf ZXZ$, we measure the middle qubit in the $\sf X$ basis and the others in the $\sf Z$ basis. Only Pauli operators with non-zero weight need to be measured, and operators can be grouped into sets where any pair of operators in the same set commute and therefore can be measured simultaneously with the same final basis transformation. One unique circuit must be run for each commuting set for each value of the variational parameters. \rsub{Each unique circuit was allocated 400 bitstring measurements to compute the expectation value of the associated observables.}

We used two different encoding schemes for VQE simulation of the LMG model. A naive encoding scheme represents each spin with a single qubit such that $N$ qubits are required to simulate $N$ coupled spins. To estimate the Hamiltonian expectation Pauli string measurements are made where all atoms are simultaneously in either the $\sf X$, $\sf Y$, or $\sf Z$ basis; after measuring in all three bases, the expectation of value of the Hamiltonian can be computed (e.g., for $N=3$, the three final basis transformations to measure the energy are $\sf XXX$, $\sf YYY$, and $\sf ZZZ$). The weights of the Pauli strings are given by Equation \ref{eq:3qubitHamiltonian}.

A second, much more efficient scheme, uses a Gray code encoding\cite{Hlatshwayo2022}.  We can simulate more spins with a reduced number of qubits by utilizing the symmetries in the Hamiltonian, mapping computational basis states to states $\ket{J, M}$ with maximal $J$ and $M = -J+2x$ for some non-negative integer $x$. With the this mapping, we require only $\lceil \log_2 (\lfloor N/2 \rfloor +1 ) \rceil$ qubits to encode a problem with $N$ spins. Thus problems with up to 15 spins can be implemented with only 3 qubits. By mapping the $\ket{J, M}$ to the computational basis via a Gray code ordering, we can ensure that all Pauli operators with non-zero weights can be measured with a number of circuits equal to the number of qubits + 1. Each grouping has exactly one qubit measured in the $\sf X$ basis while all others are measured in the $\sf Z$ basis, plus one additional grouping which measures all qubits in the $\sf Z$ basis. A table of weights for each size problem we used the Gray code coding on can be found in \crefrange{tab:3spinweights}{tab:15spinweights}. Further details about the individual spin and Gray code encoding schemes and ansatz state preparation circuits are provided in Appendices \ref{app.encoding}, \ref{app.circuits}.

\section{Experimental approach}
\label{sec.experiment}

We have implemented the VQE simulation on a neutral atom quantum computer with Cs qubits. The apparatus is similar to that used in our earlier implementation of a variational quantum algorithm\cite{Graham2022} with two significant changes. The first change is the incorporation of top-hat beam shaping on one of the qubit control beams to reduce the sensitivity of gate fidelity to optical alignment\cite{Gillen-Christandl2016}. The second change is the incorporation of atom-motion within the circuit operation which made it possible to use the top-hat beam for single-qubit control with low crosstalk,  as well as two-qubit entangling gates.

In order to implement the quantum gates in our neutral-atom quantum processor, we have introduced top-hat beam shaping in the Rydberg excitation beams, in particular to improve the $\sf CZ$  entangling gate fidelity by means of simultaneous addressing time-optimal gate protocol. In this setup during the $\sf CZ$  gate we address both atoms at the same time; however, this effort caused beam crosstalk between the control and target sites during the single-qubit rotations or the site-selective phase correction of single-atom phase shifts to achieve the canonical $\sf CZ$  gate. Thus, we have introduced mid-circuit atom transport by transporting the atoms closer to 2.5 $\mu$m to enhance the dipole-dipole interactions during the entangling gate and bring them back to 16 $\mu$m apart to implement single-qubit gates with reduced crosstalk. As a result, we have improved the two-qubit gate fidelity to 0.971(1) during the execution of quantum algorithms. A more detailed description about the atom transport is discussed in Appendix \ref{app.reconfigure}.

The experimental sequence starts by stochastic loading of single Cs atoms from a magneto-optical trap into a $4\times 4$ square array of blue detuned ``line array" bottle traps using a 825 nm trapping laser\cite{Graham2022}.
Atoms in the partially filled line array are rearranged and transferred in to a $2\times 2$ array of red-detuned tweezer traps formed using 1064-nm trapping light and crossed acousto-optic deflector (AOD) devices.  Global single-qubit gates are implemented on the array with 9.2 GHz microwaves driving the Cs clock transition between states $\ket{0}\equiv\ket{6s_{1/2},f=3,m=0}$ and $\ket{1}\equiv\ket{6s_{1/2},f=4,m=0}$ at a Rabi frequency of 64 kHz.

Local single-qubit ${\sf R}_z$ gates were implemented with 459 nm light blue detuned by 1.05 GHz relative to the transition from $\ket{6s_{1/2},f=4} \rightarrow \ket{7p_{1/2}}$  center of mass. The 459 nm light was shaped into a top-hat-like profile using a diffractive optical element and then rapidly pointed to selected qubits using crossed AODs (see Appendix \ref{app.experiment} for details). For single-qubit operations the atoms were separated by $16~\mu\rm m$ to avoid crosstalk from light spillover, as shown in Fig. \ref{fig:Schematic}. Local single-qubit ${\sf R}_x$ and ${\sf R}_y$ gates were implemented by combining local ${\sf R}_z$ rotations with global microwaves\cite{Graham2022}.

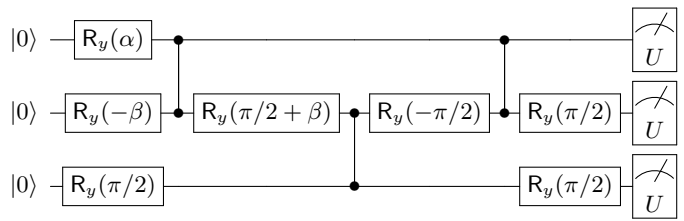
\begin{figure}[!t]
\centerline{
\Qcircuit @C=0.5em @R=.4em {
 &\lstick{\ket{0}}& \gate{{\sf R}_{y}(\alpha)} & \ctrl{1} & \qw  & \qw           &\qw            & \ctrl{1} & \qw &\meterB{U}\\
 &\lstick{\ket{0}}& \gate{{\sf R}_{y}(-\beta)} &   \ctrl{-1} &\gate{{\sf R}_{y}(\pi/2 + \beta)}  &\ctrl{1}&\gate{{\sf R}_{y}(-\pi/2)}  & \ctrl{-1}&\gate{{\sf R}_{y}(\pi/2)}  & \meterB{U}\\
 &\lstick{\ket{0}}& \gate{{\sf R}_{y}(\pi/2)} & \qw & \qw  & \ctrl{-1}                      & \qw   &\qw & \gate{{\sf R}_{y}(\pi/2)} &\meterB{U}\\
}}
\caption{Circuit for estimating the expectation value of the three-qubit Hamiltonian with individual spin encoding.
The parameters $\alpha,\beta$ are parameterized in terms of a single variational angle (see Eqs. \ref{eq:alphaparam}).
Three distinct unitary transformations are used to measure the each grouping of Pauli terms: $\sf XXX $, $\sf YYY$, and $\sf ZZZ $. These measurements are used to calculate the LMG Hamiltonian expectation value.}
\label{fig:3Q3S_circuit}
\end{figure} 

The Rydberg interaction was used to implement two-qubit $\sf CZ$ gates using a time-optimal pulse\cite{Jandura2022} which has enabled a significant recent improvement in Rydberg gate performance. The time-optimal pulse simultaneously excites both atoms with time-varying phase modulation $\phi(t)$, and has led to $\sf CZ$ fidelity above 99.3\%  with four different atomic elements: Rb\cite{Evered2023}, Sr\cite{Finkelstein2024}, Yb\cite{Peper2025,Muniz2024}, and Cs\cite{Radnaev2024}.  The Rb, Sr, and Yb results used large area Rydberg beams addressing multiple atom pairs in parallel. The Cs result employed tightly focused Rydberg beams targeting individual atoms. Here, we used an intermediate geometry with a single beam large enough to simultaneously excite a single pair of atoms to the Rydberg state. The two-photon excitation used the same 459 nm top-hat beam and laser frequency as used for the single-qubit ${\sf R}_z$ rotations together with a Gaussian profile 1040 nm beam to excite $\ket{1}$ to the $\ket{66s_{1/2},m_j=-1/2}$ Rydberg state.   

 To simultaneously excite both atoms during the gate, the targeted atoms were transported so that they were separated by $2.5~ \mu\rm m$ (see Appendix \ref{app.reconfigure}). The Rydberg beams were then directed to the point mid-way between the two atoms. Top-hat beam shaping on the 459-nm beam reduced dephasing during the gate caused by intensity variation over the atom trapping regions (see Appendix \ref{app.experiment}).  We then used local $\sf R_z$ gates to provide the proper single-qubit phase shifts for a canonical $\sf CZ$ gate. The average gate fidelity for the VQE simulations reported in the next section was ${\mathcal F}_{\sf CZ}=97.1(1)\%$ using symmetric interleaved randomized benchmarking \cite{Evered2023} as described in Appendix \ref{app.benchmarking}. Subsequent to completion of the VQE experiments, the gate parameters were further optimized to reach ${\mathcal F}_{\sf CZ}=98.6(1)\%$ (see Appendix \ref{app.bettergate} for details).

\begin{figure}[!t]
    \centering    \includegraphics[width=0.49\textwidth]{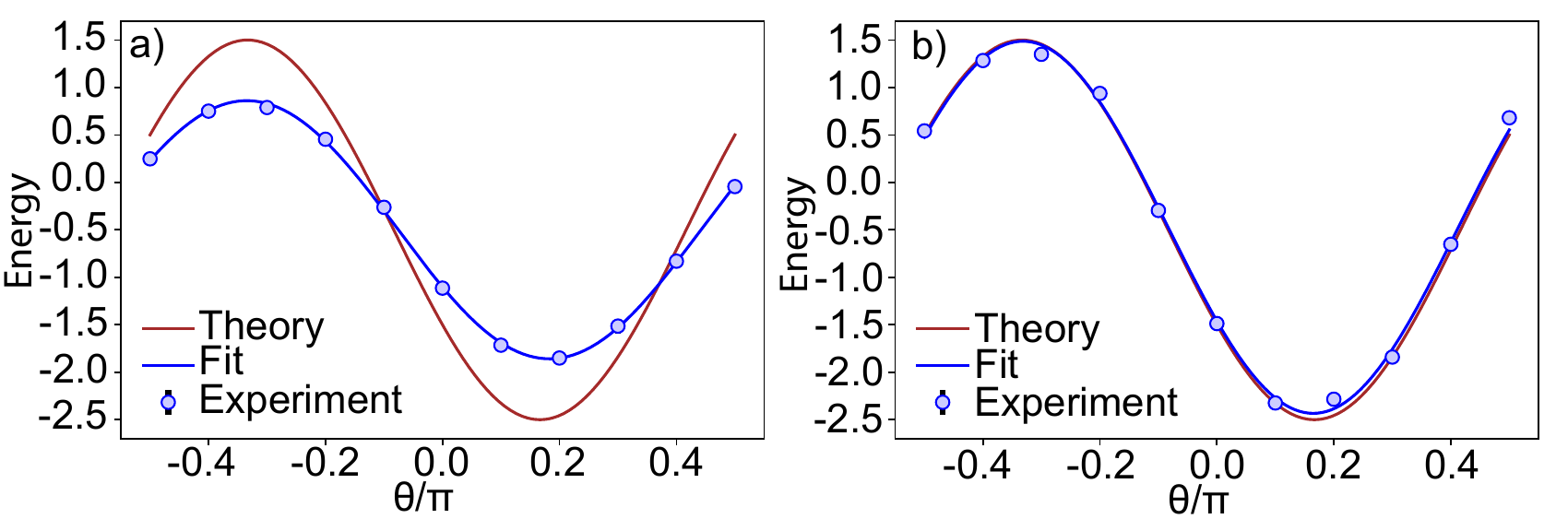}
    \caption{This figure shows the estimated $N=3$ LMG energy expectation measured using a) individual qubit encoding and b) Gray encoding. The blue and red traces in both figures are fits and theoretical curves, respectively. The energy estimates made with Gray coding show a much closer correspondence with the theoretical curve than the individual spin encoding estimates do because Gray coding more efficiently encodes spins on qubits. This greater efficiency allows all three spins to be encoding on a single qubit using Gray coding, while the individual encoding scheme requires three qubits.   In both a) and b) error bars represent the standard deviation but are smaller than the data points.}
    \label{fig:VQE_3Q_combined}
\end{figure}

\begin{table*}[!t]
    \caption{Summary of VQE simulation of the ground-state energy for $N$ spins with individual and Gray code encoding. The figure of merit is the percentage fractional difference  of the energy $PFD=\frac{|E_{\rm theory}-E_{\rm VQE}|}{|E_{\rm theory}|} \times 100. $ }
\begin{tabular}{|l|c|c|c|c|c|c|c|c}
    \hline
    &$N$ &variational  & Theory & VQE & PFD (\%)& VQE  & PFD  (\%)  \\
    & & parameters & & & &with  ZNE &with  ZNE \\
    \hline
    Individual encoding & 3 &1 & -2.5 & \rsub{-}1.86(2) & 25.6(6.0) & &  \\
     \hline
    Gray encoding & 3 &1 & -2.5 & -2.43(5) & 3(2) & & \\
 Gray encoding   & 5 &2 & -5.89 &-5.3(2) & 10(3.4) & -5.7(2) & 3.2(3.4) \\
Gray encoding&    7 &3 &  -11.26 & -10.9(4)&3(4) &  \rsub{-}11.3(4) & 0(4)  \\
Gray encoding &   9  & 4&  -18.7 &-13.8(5) & 26.2(2.7)& -17.6(8) & 5.9(4.0)  \\
    \hline
    \end{tabular}
    \label{tab:3spins}
\end{table*}

\section{Simulation of ground-state energies}
\label{sec.results}

Ground state simulation results are presented for $N=3$ spins with individual and Gray code encoding in Sec. \ref{sec.resultsa}. Gray code results for more than 3 spins are presented in Sec. \ref{sec.resultsb}.

\subsection{Three spin encodings}
\label{sec.resultsa}

As described in Section \ref{sec.VQE}, we used two different qubit encoding protocols for VQE simulations. The first of these schemes encoded each spin of the LMG simulation on an individual qubit. Using this individual qubit encoding, we used the gate set described in Sec. \ref{sec.experiment} to perform 3-qubit VQE to determine the ground-state energy of the 3-spin LMG model (see Fig. \ref{fig:3Q3S_circuit}). This circuit had one variational parameter, $\theta$, which was scanned to find the minimum energy. For each $\theta$ value, we made subsequent measurements of the qubits in the $\sf XXX, YYY$  and $\sf ZZZ$ bases. These measurements were used to compute the expectation of the LMG Hamiltonian (see Fig. \ref{fig:VQE_3Q_combined}a)
using Equation \ref{eq:3qubitHamiltonian}.
We fit the resulting energy versus $\theta$ to a cosine function to find a ground-state energy of -1.86(2) (-2.5 theoretical). The experimental error in this measurement was primarily limited by the $\sf CZ$ gate fidelity.

To improve the ground-state energy estimation, we switched to a different qubit encoding scheme based on a Gray code encoding as described in Appendix \ref{app.encoding}. This encoding technique uses qubits to encode spins much more efficiently, allowing us to encode three spins using a single qubit. 
\rsub{As is explained in detail in Appendix \ref{app.encodeGray} the ansatz state for $N$ spins has $J= N/2$ variational parameters $\theta_j$ and takes the form
\begin{align}
 \ket{\Psi_N(\Theta)} &= \sum_{k=0}^{\lfloor J \rfloor-1}
 \cos(\theta_{k+1})\prod_{l=0}^{k-1}\sin(\theta_{l+1})\ket{g_k} \nonumber\\
 &+\prod_{l=0}^{\lfloor J \rfloor -1}\sin(\theta_{l+1})\ket{g_{\lfloor J \rfloor}}.
\label{eq:GrayCodeFormula0}
\end{align}
where $\Theta=\theta_1, \theta_2,...\theta_{J} $.}

The corresponding three-spin VQE circuit is composed of single-qubit gates with the same variational parameter $\theta$ that was used in the individual spin encoding scheme.  This simplification allows improved ground-state energy estimation. In the Gray code basis, the ground-state energy can be determined by making measurements in the $\sf X $ and $ \sf Z$ bases as explained in Appendix \ref{app.encodeGray} using the weights in Table \ref{tab:3spinweights}. 
We determined the energy as a function of $\theta$ and fit to a cosine function (see Fig. \ref{fig:VQE_3Q_combined}b) to find a ground-state energy of -2.43(5) (much closer to the theoretical value of -2.5).  A comparison of these two encoding techniques is summarized in Table \ref{tab:3spins}. 

\begin{figure}[!t]
\Large
\centerline{
\Qcircuit @C=1em @R=.4em {
 &\lstick{\ket{0}}& \gate{{\sf R}_{y}}   & \ctrl{1}  &  \gate{{\sf R}_{y}}  & \meter\\
 &\lstick{\ket{0}}& \gate{{\sf R}_{y}}   & \ctrl{-1} &  \gate{{\sf R}_{y}}  & \meter\\
}}
\caption{Circuit for building generic two-qubit  states used in the five- and seven-particle Gray code encoding. The methodology used for circuit construction is described in Appendix \ref{app.circuits}.}
\label{fig:2Q_Gray_Circuit}
\end{figure}
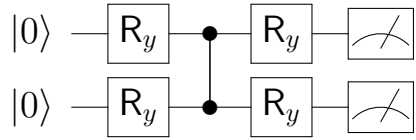

\begin{figure*}[!t]
    \centering
    \includegraphics[width=6.5in]{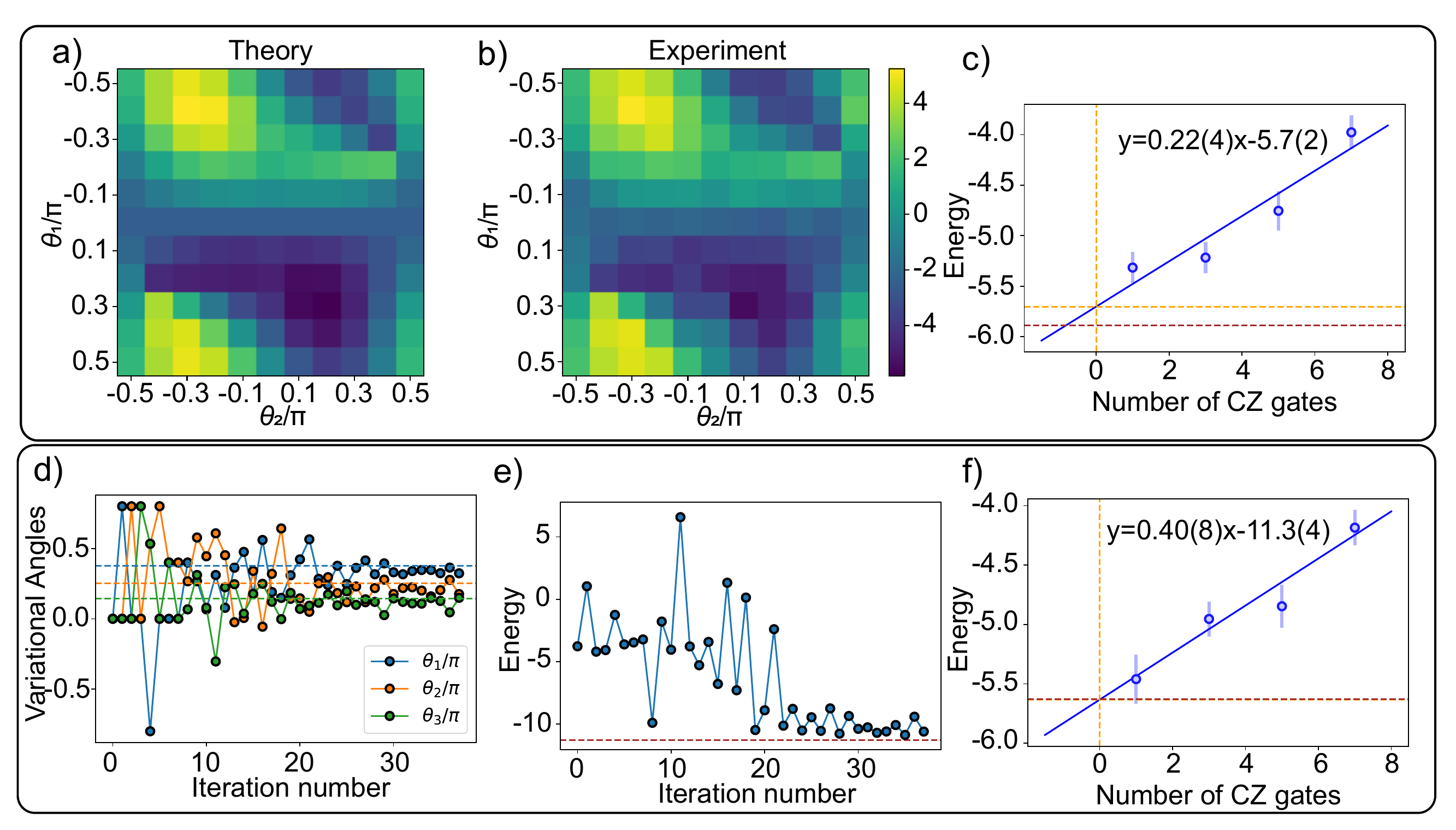}
    \caption{Results of 5-spin and 7-spin energy estimates, where the spins are encoded on 2 qubits with Gray coding. a) and b) show the theoretical and experimental energy expectation for $N=5$, respectively. The energy is plotted versus variational parameters $\theta_1$ and $\theta_2$.  c) FIIM ZNE was used to predict the $N=5$ ground-state energy for zero $\sf CZ$ noise. These measurements were taken using the variational parameters corresponding to the minimum energy found from the experimental raster scan shown in b). The solid blue line shows a linear fit of the data, the dashed orange lines intersect at the zero $\sf CZ$ noise estimate, and the brown dashed line marks the theoretical ground-state energy. d) Shows the variational parameter optimization using Nelder-Mead to minimize the measured energy expectation for the $N=7$ LMG Hamiltonian. The dashed lines represent the variational angle settings of the ideal ground state.  e) Shows the energy expectation measured during the $N=7$ optimization. The dashed brown line marks the theoretically predicted ground-state energy. f) FIIM ZNE was used to predict the $N=7$ ground-state energy for zero $\sf CZ$ noise. These measurements were taken and the variational angle settings found during the optimization shown in d) and e). The various solid and dashed lines in the figure mark the same plot features as those in defined in c). In both d) and e) error bars are smaller than the data points. All error bars represent standard deviations.}
    \label{fig:2Q_5S_7S}
\end{figure*}

%\newline
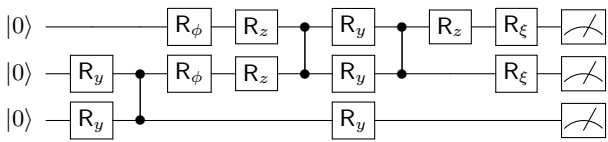
\begin{figure}[!t]
\centerline{
\Qcircuit @C=1em @R=.4em {
 &\lstick{\ket{0}}& \qw            & \qw  & \gate{{\sf R}_{\phi}} & \gate{{\sf R}_{z}} & \ctrl{1}  & \gate{{\sf R}_{y}} & \ctrl{1}  & \gate{{\sf R}_{z}} & \gate{{\sf R}_{\xi}}  & \meter\\
 &\lstick{\ket{0}}& \gate{{\sf R}_{y}}   & \ctrl{1} & \gate{{\sf R}_{\phi}} & \gate{{\sf R}_{z}} & \ctrl{-1} & \gate{{\sf R}_{y}} & \ctrl{-1} & \qw          & \gate{{\sf R}_{\xi}}  & \meter\\
 &\lstick{\ket{0}}& \gate{{\sf R}_{y}}   & \ctrl{-1}      & \qw             & \qw          & \qw       & \gate{{\sf R}_{y}} & \qw       & \qw          & \qw             & \meter\\
}}
\caption{Circuit for building generic three-qubit ansatz states used in the nine- and fifteen- particle Gray code optimization. The angles $\phi$ and $\xi$ are angles of rotation axes in the $x-y$ plane for the single qubit gates. All gates have distinct rotation angles, which are omitted for brevity and are numerically determined by decomposition code. The methodology used for circuit construction is described in Appendix \ref{app.circuits}.}
\label{fig:3Q_Gray_circuit}
\end{figure}

\subsection{More than three spins}
\label{sec.resultsb}

We used the Gray code basis to perform VQE and measure ground-state energies for 3, 5, 7, 9, and 15 spins in the LMG model. As discussed above, 3 spins can be encoded on a single qubit. Two qubits were sufficient to encode 5 and 7 spins. The same general quantum circuit (containing one $\sf CZ$ gate) could be used to perform the VQE algorithm for both 5 and 7 spins (see Fig. \ref{fig:2Q_Gray_Circuit}). Similarly, for both spins, the LMG energy expectation value could be determined from $\sf ZZ $, $\sf XZ $ and $ \sf ZX$ Pauli string measurements
using the weights in Tables \ref{tab:5spinweights} and \ref{tab:7spinweights}.
However, the 5 and 7 spin ansatz states have a different number of variational parameters (2 and 3 parameters, respectively). This difference led us to use two different strategies to determine the optimal parameter settings.  For 5 spins, we performed a 2D raster scan of parameters to find the values that gave the minimum energy (see Fig. \ref{fig:2Q_5S_7S}a and b). This strategy would have been much slower to optimize the 3 parameters of the 7-spin ansatz state, so we used the Nelder-Mead algorithm to optimize the variational parameters to give the minimum energy (see Fig. \ref{fig:2Q_5S_7S}d and e). 

Once the optimal variational parameter settings were determined, we used zero-noise extrapolation with fixed identity insertions (FIIM ZNE) \cite{Dumitrescu2018,Pascuzzi2022} to reduce the impact of  $\sf CZ$ gate error on the ground-state energy estimates. In this technique, the fact that two subsequent $\sf CZ$ gates are ideally equivalent to the identity operator is used to increase the noise due to the $\sf CZ$ gates in a controlled fashion. By adding a varying number of identity insertions (pairs of $\sf CZ$ gates) at each $\sf CZ$ gate in the circuit, the energy as a function of the $\sf CZ$ number can then be used to linearly extrapolate what the energy would be if there were zero $\sf CZ$ gates. We used FIIM ZNE with 1 and 2 identity insertions (2 and 4 additional $\sf CZ$ gates, respectively) to linearly extrapolate to energy estimates with zero $\sf CZ$ gate error (see Fig. \ref{fig:2Q_5S_7S}c and f). More details about ZNE are found in Appendix \ref{app.ZNE}. Using FIIM ZNE, we estimated the LMG Hamiltonian ground-state energy for 5 and 7 spins to be -5.7(2) (-5.89 theoretical) and -11.3(4) (-11.26 theoretical).

\begin{figure}[!t]
    \centering
    \includegraphics[width=3.3in]{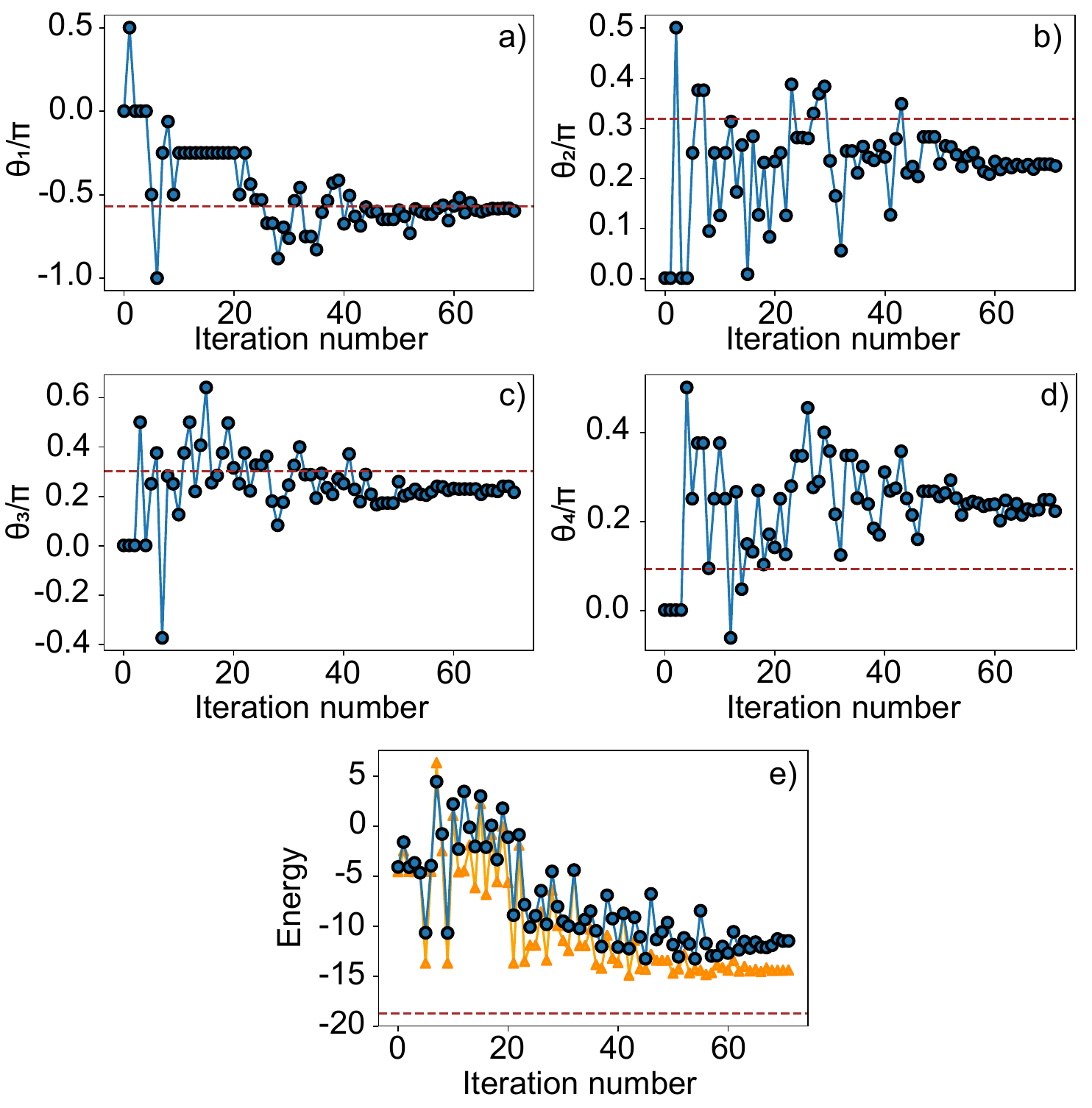}
    \caption{Nelder Mead optimization of the $N=9$ ansatz state to minimize the LMG energy. a-d) show the variational parameters ($\theta_1$-$\theta_4$) during the optimization. The dashed brown lines mark the parameter values of the ground state. e) shows the measured energy expectation during the optimization. The orange and teal points show the theoretical and experimentally measured energy for the parameter set used in each iteration, and the dashed brown line marks the ground-state energy. This optimization did not converge to the ideal ground-state parameter set. The optimized parameter set was refined using linear scans of each parameter (see Fig. \ref{fig:9spinslinearscans}).}
    \label{fig:9spins}
\end{figure}

\begin{figure}[!t]
    \centering
    \includegraphics[width=2.7in]{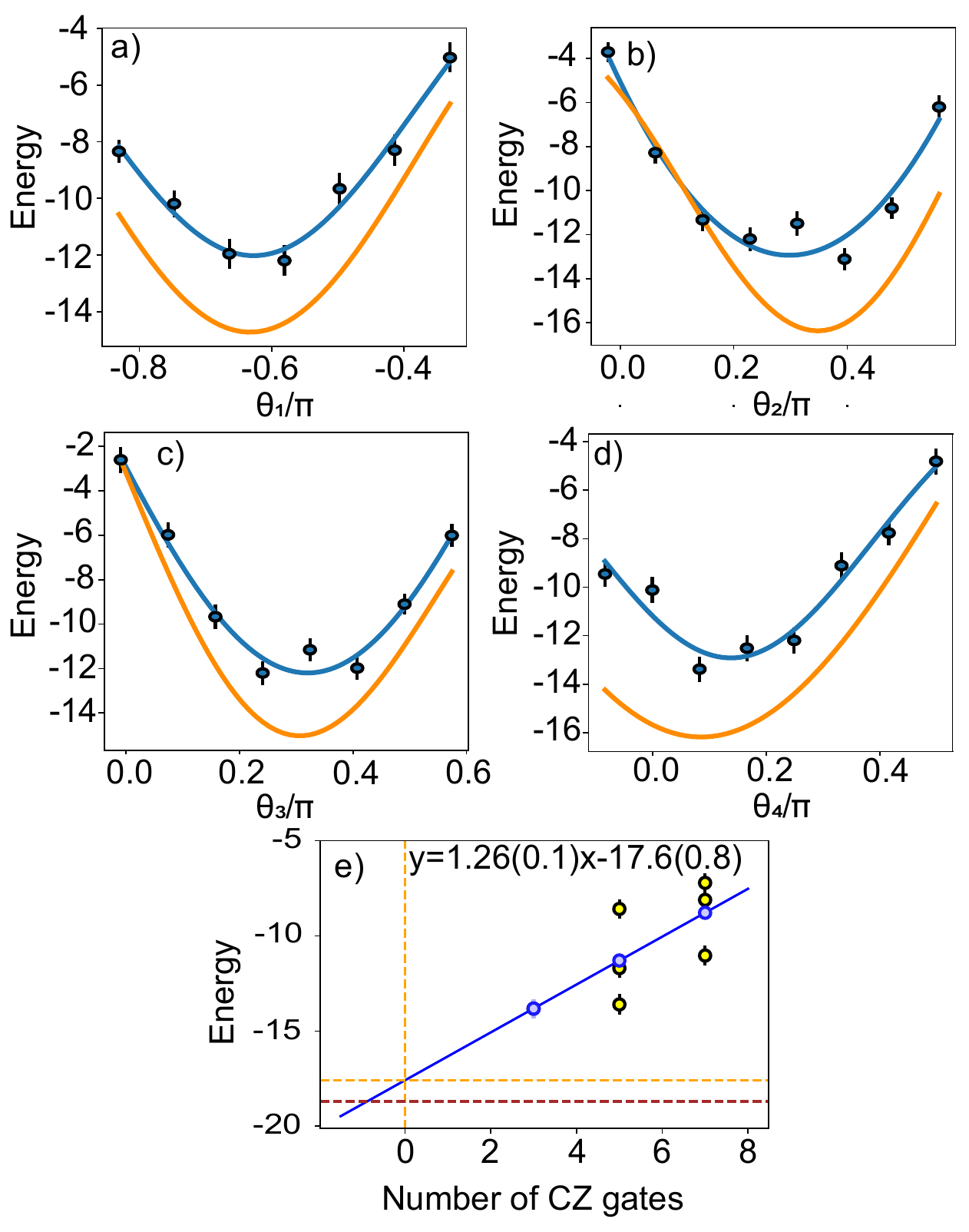}
    \caption{Panels a-d) show linear scans of the $N=9$ ansatz state parameters. Blue and orange traces correspond to a fit to the measured values  and the theoretical energy. e) SIIM ZNE was used to predict the $N=9$ ground-state energy for zero $\sf CZ$ noise. These measurements were taken using the variational parameters corresponding to the minimum energy found from the scans shown in a-d). Yellow data points mark the measured energy for $\sf CZ$-pair identity insertions made at each of the three $\sf CZ$ gates in the circuit. Blue data points mark the average of the yellow points made at that $\sf CZ$ number. The solid blue line shows a linear fit of the data, the dashed orange lines intersect at the zero $\sf CZ$ noise estimate, and the brown dashed line marks the theoretical ground-state energy.}
    \label{fig:9spinslinearscans}
\end{figure}

Three qubits were needed to encode the ansatz states for 9 and 15 spins. All three-qubit VQE algorithms use the same general quantum circuit with three $\sf CZ$ gates (see Fig. \ref{fig:3Q_Gray_circuit}). Both spin numbers also require the same Pauli string measurements ($\sf ZZZ $, $\sf XZZ $, $\sf ZXZ$, and $\sf ZZX$) to determine the energy expectation of the LMG Hamiltonian
using the weights in Tables \ref{tab:9spinweights} and \ref{tab:15spinweights}.
The ansatz states for 9 and 15 spins required 4 and 7 variational parameters, respectively, so we used the Nelder-Mead algorithm to find the optimal variational parameter settings, as we did for the 7-spin VQE optimization. The optimization protocol did not fully converge to the optimal variational parameter settings for either spin number. The 9-spin optimization converged to a variational parameter set which corresponded to a theoretical energy expectation 23\% higher than the theoretical ground-state energy (see Fig. \ref{fig:9spins}a-e). We further refined the 9-spin variational parameter settings with 1D line scans of each of the four angles and fit the energies of each set to a parabola (see Fig. \ref{fig:9spinslinearscans}). The set of variational parameters from these fits corresponded to a theoretical energy value within 3.3\% of the ground-state energy. The larger number of parameters  and more complicated optimization landscape prevented the 15-spin VQE algorithm from converging to a variational parameter set near the theoretical optimum (the theoretical value of the converged parameter set was 48\% higher than the theoretical ground state energy); this large error prevented us from finding an accurate ground-state energy estimate for 15 spins. Further discussion of the 15-spin VQE optimization is found in Appendix \ref{app.15spinVQE}. 

Since the 9-spin VQE circuit required 3 $\sf CZ$ gates, the FIIM ZNE method introduced above would require 6 additional CZ gates for each identity insertion, and one and two identity insertions would require circuits with 9 and 15 $\sf CZ$ gates, respectively. Circuits with this number of additional $\sf CZ$ gates would yield energy values outside the linear noise regime, preventing linear extrapolation to the zero noise condition. Instead, we used an alternative technique in which identity insertions need not be made simultaneously at every $\sf CZ$ gate in the circuit. We made identity insertions on a single $\sf CZ$ gate in the circuit at a time. This single identity insertion is repeated for the other two $\sf CZ$ gates in the circuit. The results of the circuits with the same number of $\sf CZ$ gates are then averaged. This process can then be repeated for other numbers of identity insertions. The averages are then used to extrapolate to zero $\sf CZ$ error. This zero-noise extrapolation technique is known as the ``set identity insertion method" (SIIM) \cite{Pascuzzi2022}. We used this technique with 1 and 2 identity insertions made at each of the three $\sf CZ$ gates in the circuit (see Fig. \ref{fig:9spinslinearscans}e) and estimated the ground-state energy to be -17.6(8) (-18.7 theoretical). These results are summarized in Table \ref{tab:3spins}.

\section{Summary}
\label{sec.summary}

We have demonstrated simulation of ground-state energies of the LMG Hamiltonian using the VQE algorithm for up to 15 interacting spins using two different encoding schemes. The closed-loop VQE implementation used a Nelder-Mead algorithm for the classical optimization of the ansatz state parameters. 
 The individual encoding scheme requires a number of qubits equal to the number of spins, but only three separate Pauli string measurements to extract the energy. 
 The Gray encoding is exponentially more efficient in the number of qubits required at the cost of an increased number of variational parameters and an increased number of Pauli string measurements. 

 The agreement of the simulated ground-state energies with exact LMG model results was improved by using ZNE for error mitigation. Error mitigation with ZNE provided close agreement between the VQE results and theory for up to 7 spins and agreement to about 5\% for 9 spins. Attempts to extend to a 15-spin problem with 7 variational parameters resulted in poor convergence. \rsub{As is shown in Appendix \ref{app.15spinVQE}  the lack of convergence is  due not just to gate execution errors, but  also results from  the possible existence of barren  plateaus in the high-dimensional state space and  sensitivity to shot-noise  from the limited number of bitstring measurements allocated to each circuit. These factors resulted in the Nelder-Mead optimizer getting trapped in local minima without finding the global optimal solution.} 

\rsub{Possible solutions may involve an expanded  
search over multiple initial conditions together with dynamic allocation of the number of iterations used for each circuit.} To reduce the impact of gate errors full quantum error correction is ultimately the goal, but this will require many more qubits and higher $\sf CZ$ fidelity than was available here. 
Alternatively,  it will be of interest to consider an intermediate approach that goes beyond error mitigation with small error detecting codes. Such codes can provide improved results with only a moderate increase in qubit overhead, as has recently been demonstrated for quantum chemistry and quantum materials simulations\cite{vanDam2024,Bedalov2024}.

\section{Acknowledgements}
This material is based on work supported by the
U.S. Department
of Energy, Office of Science, Office of High Energy Physics, under Award No. DE-SC0019465, 
the U.S. Department
of Energy under Argonne contract 2F-60009,
the U. S.  Department of Energy Office of Science National Quantum Information Science Research Centers as
part of the Q-NEXT center, 
as well as support from NSF Award 2016136 for the QLCI center Hybrid
Quantum Architectures and Networks, 
NSF award PHY-2411495 
and Infleqtion, Inc.  

\appendix 

\section{Encoding schemes and ansatz states}
\label{app.encoding}

\subsection{Individual spin basis}

The most direct choice for encoding the LMG model onto a quantum computer is to have the state of each qubit in the register correspond to the spin state of one particle. Thus, the number of qubits is equal to the number of particles. In this basis, the Hamiltonian for $N$ particles is given by
\begin{equation}
\label{eq:spinH}
H = \sum_{p=1}^N \frac{1}{2}{\sf Z}^{(p)} + \frac{V}{2}  \sum_{\substack{p,q=1\\q\neq p}}^N \left( {\sf X}^{(p)} {\sf X}^{(q)} - {\sf Y}^{(q)} {\sf Y}^{(p)} \right) .
\end{equation}
We can form three groups of operators from these sums containing the $\braket{{\sf Z^{p}}}$, $\braket{{\sf X^{p}X^{q}}}$, and $\braket{{\sf Y^{p}Y^{q}}}$ terms by measuring in the $\sf ZZZ$, $\sf XXX$, and $\sf YYY$ bases, respectively. For the $\sf ZZZ$ basis, no basis transformation is necessary. For the $\sf XXX$ basis, the basis is transformed by a global ${\sf R}_y(-\pi/2)$ gate. For the $\sf YYY$ basis, the transformation is done by a global ${\sf R}_x(\pi/2)$ gate. Thus, for each specific value of the variational parameters to evaluate, three distinct circuits must be executed several times to obtain an estimate of the energy. 

We choose a trial state by first solving for the ground state of the Hamiltonian as a function of the interaction parameter $V$. For the LMG model, this may be done exactly and allows for comparison of experimental results with the exact solution. Then, we rewrite the solution in terms of trigonometric functions of variational angles. For some specific choice of $V$, say $V=1$, there is a corresponding value of the variational angles such that the trial wave function is the true ground state of the system. The true ground states have contributions from the computational basis states whose Hamming weights have the same parity as the number of particles in the associated problem. 

As an example, we describe the procedure for $N=3$. 
Using the computational basis 
$ \{\ket{000},\ket{001},\ket{010},\ket{011},\cdots,\ket{111}\}$ 
the Hamiltonian is represented by 
\begin{eqnarray}
H &=& \frac{1}{2}( {\sf Z} \otimes {\sf I} \otimes {\sf I}  + {\sf I} \otimes {\sf Z} \otimes {\sf I} + {\sf I} \otimes {\sf I} \otimes {\sf Z} ) \nonumber \\ 
 &+& \frac{V}{2} ({\sf I}  \otimes {\sf X} \otimes {\sf X}  + {\sf X}  \otimes {\sf X}  \otimes {\sf I}  + {\sf X}  \otimes {\sf I} \otimes {\sf X} \nonumber \\
&-& {\sf I}  \otimes {\sf Y}  \otimes {\sf Y}  - {\sf Y}  \otimes{\sf Y}  \otimes {\sf I}  - {\sf Y}  \otimes {\sf I}  \otimes {\sf Y} ).
\label{eq:3qubitHamiltonian}
\end{eqnarray}
an appropriate variational ansatz  for the ground state of $N=3$  is 
\begin{align}
 \ket{\psi(\theta)} = &\cos(\theta)\ket{111} 
- \frac{1}{\sqrt{3}} \sin(\theta) \big(\! \ket{001} + \ket{010} + \ket{100} \!\big),
\label{eq:N=3trial}
\end{align}
where $\theta$ is the variational parameter.  This wave function is the ground state when $\sqrt{3}\cot(\theta)=V/(1+\sqrt{1+3V^2})$ with energy $-1/2-\sqrt{1+3V^2}$.
The three-qubit preparation circuit  is shown in Fig.~\ref{fig:3Q3S_circuit}, written with two auxiliary angles $\alpha$ and $\beta$ defined by
\begin{subequations}
\begin{eqnarray}
\alpha &=& 2\arccos\bigg(-\sqrt{\frac{2}{3}}\sin\theta\bigg),  \\
\beta &=& -\frac{\pi}{4} - \arctan\bigg(\frac{\tan\theta}{\sqrt{3}}\bigg). \label{eq:betaparam}
\end{eqnarray}
\label{eq:alphaparam}
\end{subequations}

\subsection{Gray code basis}
\label{app.encodeGray}

Each of the trial states are superpositions that have contributions from the computational basis states whose Hamming weights have the same parity as the number of particles in the associated problem. Additionally, each of the computational basis states with the same Hamming weight shares the same probability amplitude. These properties are direct consequences of the ground state of the LMG model having maximal $J$. The qubit basis is able to access states with arbitrary $J$, so it is not necessarily the most efficient use of quantum state space. We can alternatively express the trial states in the $\ket{J,M}$ basis for $N=2,3,4$ corresponding to $J=1,3/2,2$ as
\begin{subequations}
\begin{eqnarray}
    \ket{\Psi_2} &=&  \cos\theta\ket{1, -1}-\sin\theta\ket{1, 1},\\
    \ket{\Psi_3} &=& \cos\theta\ket{3/2, -3/2}-\sin\theta\ket{3/2, 1/2},\\
    \ket{\Psi_4} &=& \cos^2\theta \ket{2, -2} + \sin^2\theta \ket{2, 2}  -\frac{1}{\sqrt{2}} \sin2\theta \ket{2, 0}  .\nonumber\\
\end{eqnarray}
\label{eq:jmjbasisn}
\end{subequations}
When expressed in the $\ket{J,M}$ basis, the trial states require fewer states than the individual spin basis. In fact, the LMG Hamiltonian of Eq. \eqref{eq:spinH} only couples states with $\Delta M = \pm 2$, and therefore at most $d=\lfloor J \rfloor +1$ states have non-zero probability amplitudes. This suggests that a more appropriate encoding directly maps computational basis states to the $\ket{J, M}$ basis, requiring at most $\lceil \log_2 (\lfloor N/2 \rfloor +1) \rceil$ qubits to encode a problem of $N$ particles. By mapping these states with a Gray code ordering, we can minimize the number of Pauli string groupings and therefore the number of unique circuits to run\cite{Hlatshwayo2022, DiMatteo2021}.

A Gray code is an ordering of binary values where any two adjacent entries in the code differ by only a single bit. For example, an ordering of two bit values may be given by 
\begin{equation}
    \label{eq:gray_example}
    00 \rightarrow 0\mathbf{1} \rightarrow \mathbf{1}1 \rightarrow 1\mathbf{0}.
\end{equation}
There are multiple possible Gray codes $\mathbf{G}_\nu$ on $\nu$ bits, each defined by 
\begin{equation}
    \label{eq:gray_definition}
    \mathbf{G}_\nu = (g_0, g_1, \ldots,  g_{2^\nu-1}).
\end{equation}.

A binary reflective Gray code on $\nu$ bits can be generated recursively in the following manner. Let $\overline{\mathbf{G}_\nu}$ represent a Gray code where the entries $g_\alpha$ appear in the same order but with their bits reversed. Then the binary reflected Gray code is given by\cite{DiMatteo2021} 
\begin{equation}
    \label{eq:brgc_def}
    \mathbf{G}_\nu = (\mathbf{G}_{\nu-1}\cdot 0,\overline{\mathbf{G}_{\nu-1}}\cdot 1),
\end{equation}
where the center dot indicates concatenation. With this definition, the LMG Hamiltonian may be written as\cite{Hlatshwayo2022}
\begin{equation}
    \label{eq:GrayH}
    H = \sum_{k=0}^{d-1} a_k \ket{g_k}\bra{g_k} +\sum_{k=0}^{d-2} b_k [\ket{g_k}\bra{g_{k+1}}+\ket{g_{k+1}}\bra{g_k}],
\end{equation}
where the coefficients are given by 
\begin{subequations}
\begin{eqnarray}
    \label{eq:gray_diag_H}
    a_k &=& \epsilon [2k-J] = \epsilon M,\\
    \label{eq:gray_off_diag_H}
    b_k &=& -\frac{V}{2} F(M = 2k-J),\\
    F(M) &=& \{[J(J+1)-M(M+1)] \\ &\times& [J(J+1)-(M+1)(M+2)]\}^{\frac{1}{2}} .
\end{eqnarray}
\end{subequations}
With this choice of encoding, the trial states can be expressed in terms of the computational basis states as 
\begin{subequations}
\begin{eqnarray}
    \ket{\Psi_2} &=& \cos(\theta)\ket{0}-\sin(\theta)\ket{1}
\label{eq:graybasisn=2}\\
    \ket{\Psi_3} &=& \cos(\theta)\ket{0}-\sin(\theta)\ket{1}
\label{eq:graybasisn=3}
\\
    \ket{\Psi_4} &=& \cos^2(\theta) \ket{00} + \sin^2(\theta) \ket{11} 
 -\frac{1}{\sqrt{2}} \sin(2\theta) \ket{01}  \nonumber \\
\label{eq:graybasisn=4}
\end{eqnarray}
\end{subequations}
The $N=2$- and $N=3$-particle cases reduce to a single-qubit state, and the four-particle case reduces to two qubits. The number of qubits used is logarithmic in the number of particles, and the number of $\sf CNOT$ gates to prepare quantum states scales exponentially in the number of qubits, leading to an overall linear scaling in the number of $\sf CNOT$ gates required to prepare an N-particle trial state. In order to guarantee that the true ground state is obtained by some values of the variational parameters, we perform a search over all possible states with real-valued probability amplitudes. To cover arbitrary real superpositions of the $d$ basis states, $d-1 = \lfloor N / 2 \rfloor $ variational angles are necessary. In the special case of $N=4$, we take advantage of the fact that two amplitudes in the ground state always sum to one to reduce the number of angles needed.

One advantage of the individual qubit encoding for the LMG Hamiltonian is that regardless of the number of particles in the problem, the energy expectation value can always be determined by measuring every qubit in one of three bases. In the Gray code encoding, there is an alternative pattern for the necessary bases to measure energies. The Gray code ordering ensures that interactions occur only between states that differ by at most one bit. Additionally, the Hamiltonian is real valued. Consequently, the non-zero terms contain no $\sf Y$ basis measurements on any qubits, and contain an $\sf X$ basis measurement on at most one qubit at a time. Therefore, all non-zero terms for a $\nu$-qubit Hamiltonian can be evaluated with $\nu+1$ basis measurements: one where all qubits are measured in the $\sf Z$ basis (unperturbed Hamiltonian), and one where a single qubit is measured in the $\sf X$ basis and the remaining qubits in the $\sf Z$ basis for each qubit\cite{DiMatteo2021}. The number of required basis measurements increases with qubit count and is therefore logarithmic in particle number. The estimate of the energy is derived from a weighted sum of the expectation value of the Pauli strings $P_i$: $\bra{\psi} H \ket{\psi} = \sum_{i} a_i \bra{\psi} P_i \ket{\psi}$. The weights are given by $a_i = {\rm Tr}(HP_i^{\dagger})/{\rm dim}(H)$. Weights corresponding to each of the contributing Pauli strings for each size problem are listed in Tables \ref{tab:3spinweights} to \ref{tab:15spinweights}.

When there is a sufficiently small number of particles in the problem, it is feasible to solve for probability amplitudes that are simple trigonometric functions of an angle that itself depends on the interaction strength $V$ and for which an appropriate choice of the variational angle $\theta$ will produce the ground state for any interaction strength $V$. When enough particles are added into the problem, finding algebraic solutions for this construction becomes infeasible, and we instead opt for an ansatz that respects the known symmetries of the solution space. The possible non-zero probability amplitudes are the states that satisfy $M_J = -J+2x$ for some non-negative integer $x$. Each of the probability amplitudes is real-valued. We thus defined our ansatz states to cover the space of arbitrary real-valued states with the multiplicity of non-zero probability amplitudes determined by the particle number.

The ansatz states with Gray encoding for $N=5,7,9,15$ spins are
\begin{align}
 \ket{\Psi_5(\theta_1, \theta_2)} &=
 \cos(\theta_1)\ket{00} \nonumber\\
 &+ \sin(\theta_1)\cos(\theta_2) \ket{01} \nonumber \\
 &+ \sin(\theta_1)\sin(\theta_2) \ket{11} ,
\label{eq:N=5trialGray}
\end{align}
\begin{align}
 \ket{\Psi_7(\theta_1, \theta_2, \theta_3)} &= \cos(\theta_1)\ket{00} \nonumber\\
 &+ \sin(\theta_1)\cos(\theta_2) \ket{01} \nonumber \\
 &+ \sin(\theta_1)\sin(\theta_2)\cos(\theta_3) \ket{11} \nonumber\\
 &+ \sin(\theta_1)\sin(\theta_2)\sin(\theta_3) \ket{10}, 
\label{eq:N=7trialGray}
\end{align}

\begin{align}
 \ket{\Psi_9 (\theta_1, \theta_2, \theta_3,\theta_4)} &=  \cos(\theta_1)\ket{000}\nonumber\\
 &+ \sin(\theta_1)\cos(\theta_2) \ket{001} \nonumber \\
  &+ \sin(\theta_1)\sin(\theta_2)\cos(\theta_3) \ket{011} \nonumber \\
  &+ \sin(\theta_1)\sin(\theta_2)\sin(\theta_3) \cos(\theta_4) \ket{010} \nonumber \\
  &+ \sin(\theta_1)\sin(\theta_2)\sin(\theta_3) \sin(\theta_4)\ket{110},  
\label{eq:N=9trialGray}
\end{align}

\begin{align}
 &\ket{\Psi_{15}(\theta_1, \theta_2, \theta_3,\theta_4, \theta_5, \theta_6, \theta_7)}=\nonumber\\ 
&   \cos(\theta_1)\ket{000}   \nonumber \\
&+\sin(\theta_1)\cos(\theta_2) \ket{001}\nonumber\\
  &+ \sin(\theta_1)\sin(\theta_2)\cos(\theta_3) \ket{011} \nonumber \\
  &+ \sin(\theta_1)\sin(\theta_2)\sin(\theta_3) \cos(\theta_4) \ket{010} \nonumber \\
  &+ \sin(\theta_1)\sin(\theta_2)\sin(\theta_3) \sin(\theta_4)\cos(\theta_5)\ket{110} \nonumber \\ 
    &+ \sin(\theta_1)\sin(\theta_2)\sin(\theta_3) \sin(\theta_4)\sin(\theta_5)\cos(\theta_6)\ket{111} \nonumber \\ 
    &+ \sin(\theta_1)\sin(\theta_2)\sin(\theta_3) \sin(\theta_4)\sin(\theta_5)\sin(\theta_6)\cos(\theta_7)\ket{101} \nonumber \\ 
    &+ \sin(\theta_1)\sin(\theta_2)\sin(\theta_3) \sin(\theta_4)\sin(\theta_5)\sin(\theta_6)\sin(\theta_7)\ket{100} 
\label{eq:N=15trialGray}
\end{align}

The general formula for arbitrary $N\rsub{\ge 2}$ is
\begin{align}
 \ket{\Psi_N(\Theta)} &= \sum_{k=0}^{\lfloor J \rfloor-1}
 \cos(\theta_{k+1})\prod_{l=0}^{k-1}\sin(\theta_{l+1})\ket{g_k} \nonumber\\
 &+\prod_{l=0}^{\lfloor J \rfloor \rsub{-1}}\sin(\theta_{l+1})\ket{g_{\lfloor J \rfloor}}
\label{eq:GrayCodeFormula}
\end{align}
\rsub{with $J= N/2$ and $\Theta=\theta_1, \theta_2,...\theta_{J} $.}

\setcounter{table}{0}
\renewcommand{\thetable}{A\arabic{table}}

\begin{table*}[!t]
    \caption{Weights of the Pauli string decomposition for the $N=3$ Gray code encoded LMG Hamiltonian.}

    \begin{tabular}{|c|c|c|c|c|c|c|}
    \hline
    \multicolumn{2}{|c|}{{\sf Z} Grouping} & \multicolumn{2}{|c|}{{\sf X} Grouping}\\
    \hline
    Pauli String  & Weight& Pauli String & Weight\\
    \hline
     {\sf I} &$-1/2$ & {\sf X} &$-\sqrt{3} V$ \\
     {\sf Z} &$-1$  &  &\\

    \hline
    \end{tabular}
    \label{tab:3spinweights}
\end{table*}
\begin{table*}[!t]
    \caption{Weights for the $N=5$ spin Gray code encoded LMG Hamiltonian.}

    \begin{tabular}{|c|c|c|c|c|c|c|c|c}
    \hline
    \multicolumn{2}{|c|}{{\sf ZZ} Grouping} & \multicolumn{2}{|c|}{{\sf XZ} Grouping} & \multicolumn{2}{|c|}{{\sf ZX} Grouping}\\
    \hline
    Pauli String  & Weight& Pauli String & Weight & Pauli String & Weight\\
    \hline
    {\sf  II} &$-\frac{3}{8}$ & {\sf XI} &$-\frac{3 V}{\sqrt{2}}$& {\sf IX} &$-\sqrt{\frac{5}{2}}V$ \\
     {\sf IZ} &$-\frac{7}{8}$  & {\sf XZ} &$\frac{3 V}{\sqrt{2}}$& {\sf ZX} &$-\sqrt{\frac{5}{2}}V$\\
     {\sf ZI} &$-\frac{9}{8}$  &  & & & \\
    \hline
    \end{tabular}
    \label{tab:5spinweights}
    \end{table*}

\begin{table*}[!t]
    \caption{Weights for the $N=7$ spin Gray code encoded LMG Hamiltonian.}

    \begin{tabular}{|c|c|c|c|c|c|c|c|c}
    \hline
    \multicolumn{2}{|c|}{{\sf ZZ} Grouping} & \multicolumn{2}{|c|}{{\sf XZ} Grouping} & \multicolumn{2}{|c|}{{\sf ZX} Grouping}\\
    \hline
    Pauli String  & Weight& Pauli String & Weight & Pauli String & Weight\\
    \hline
     {\sf II} &$-\frac{1}{2}$ & {\sf XI} &$-\sqrt{15}V$& {\sf IX} &$-\big(3\sqrt{5}+\sqrt{21}\big)\frac{V}{2}$ \\
     {\sf ZZ} &$-1$  & {\sf XZ} &$\sqrt{15}V$& {\sf ZX} &$\big(3\sqrt{5}-\sqrt{21}\big)\frac{V}{2}$\\
     {\sf ZI} &$-2$  &  & & & \\
    \hline
    \end{tabular}
    \label{tab:7spinweights}
    \end{table*}

\begin{table*}[!t]
    \caption{Weights for the $N=9$ spin Gray code encoded LMG Hamiltonian.}

    \begin{tabular}{|c|c|c|c|c|c|c|c|c|c|c|}
    \hline
    \multicolumn{2}{|c|}{{\sf ZZZ} Grouping} & \multicolumn{2}{|c|}{{\sf XZZ} Grouping} & \multicolumn{2}{|c|}{{\sf ZXZ} Grouping} & \multicolumn{2}{|c|}{{\sf ZZX} Grouping}\\
    \hline
    Pauli String  & Weight& Pauli String & Weight & Pauli String & Weight& Pauli String & Weight\\
    \hline
     III &$-\frac{5}{16}$ & {\sf XZZ} &$\frac{\sqrt{21}}{2}V$& {\sf ZXZ} &$\frac{3}{2}\sqrt{\frac{7}{2}}V$&{\sf ZZX}&$\big(5\sqrt{6}-6\big)\frac{V}{4}$ \\
     {\sf IIZ} &$\frac{7}{16}$  & {\sf XZI} &$\frac{\sqrt{21}}{2}V$& {\sf ZXI} &-$\frac{3}{2}\sqrt{\frac{7}{2}}V$&{\sf ZIX}&$-\big(5\sqrt{6}+6\big)\frac{V}{4}$\\
     {\sf IZI} &$-\frac{23}{16}$ & {\sf XIZ} &$-\frac{\sqrt{21}}{2}V$  &{\sf IXZ} &$\frac{3}{2}\sqrt{\frac{7}{2}}V$ &{\sf IZX} &$\big(5\sqrt{6}-6\big)\frac{V}{4}$\\
     {\sf IZZ} &$-\frac{5}{16}$  &  {\sf XII}&-$\frac{\sqrt{21}}{2}V$ &{\sf IXI} &$-\frac{3}{2}\sqrt{\frac{7}{2}}V$ &{\sf IIX} &$-\big(5\sqrt{6}+6\big)\frac{V}{4}$\\
     {\sf ZII} &$-\frac{19}{16}$  &  & & & &&\\
     {\sf ZIZ} &$-\frac{7}{16}$  &  & & & &&\\
     {\sf ZZI} &$-\frac{9}{16}$  &  & & & &&\\
     {\sf ZZZ} &$-\frac{1}{16}$  &  & & & &&\\
    \hline
    \end{tabular}
    \label{tab:9spinweights}
    \end{table*}

    \begin{table*}[!t]
    \caption{Weights for the $N=15$ spin Gray code encoded LMG Hamiltonian.}
    
        \begin{tabular}{|c|c|c|c|c|c|c|c|c|c|c|}
    \hline
    \multicolumn{2}{|c|}{{\sf ZZZ} Grouping} & \multicolumn{2}{|c|}{{\sf XZZ} Grouping} & \multicolumn{2}{|c|}{{\sf ZXZ} Grouping} & \multicolumn{2}{|c|}{{\sf ZZX} Grouping}\\
    \hline
    Pauli String  & Weight& Pauli String & Weight & Pauli String & Weight& Pauli String & Weight\\
    \hline
     {\sf III} &$-\frac{1}{2}$ & {\sf XZZ} &$3\sqrt{7}V$& {\sf ZXZ} &$\big(3\sqrt{13}-\sqrt{165} \big)\frac{V}{2}$&{\sf ZZX}&$\big(5\sqrt{33}-4\sqrt{105}+\sqrt{273}\big)\frac{V}{4}$ \\
     {\sf ZZI} &-2  & {\sf XZI} &$3\sqrt{7}V$& {\sf ZXI} &$\big(-3\sqrt{13}+\sqrt{165} \big)\frac{V}{2}$&{\sf ZIX}&$\big(-5\sqrt{33}+2\sqrt{105}+\sqrt{273}\big)\frac{V}{4}$\\
     {\sf ZZZ} &-1 & {\sf XIZ} &$-3\sqrt{7}V$ &{\sf IXZ} &$\big(3\sqrt{13}+\sqrt{165} \big)\frac{V}{2}$ &{\sf IZX} &$\big(5\sqrt{33}+2\sqrt{105}-\sqrt{273}\big)\frac{V}{4}$\\
      &  &  {\sf XII}&$-3\sqrt{7}V$&{\sf IXI} &$-\big(3\sqrt{13}+\sqrt{165} \big)\frac{V}{2}$ &{\sf IIX} &$\big(5\sqrt{33}+4\sqrt{105}+\sqrt{273}\big)\frac{V}{4}$\\
    \hline
    \end{tabular}
    \label{tab:15spinweights}
\end{table*}

\section{Circuits for Gray code encoding}
\label{app.circuits}

The Ansatz states used in the Gray code encoding scheme are arbitrary real-valued quantum states with specific probability amplitudes allowed to be non-zero. We follow the scheme of \cite{Plesch2011}, leveraging the Schmidt decomposition of the target Ansatz state to construct unitary operations that will prepare it. To begin, we divide the qubits into two groups $A$ and $B$. We then rewrite the quantum state as a matrix $M$ where rows refer to the states of qubits in $A$ and columns refer to the states of qubits in $B$. After applying the singular value decomposition, we have $M = U \Sigma V^{\dagger}$. The diagonal elements of $\Sigma$ provide a quantum state of subsystem $A$ such that the sequence of $\sf CNOT$ to entangle subsystems $A$ and $B$ followed by unitary operation $U$ on subsystem $A$ and $V^{\dagger}$ on subsystem $B$ will prepare the desired quantum state. Because our target final state is real valued, the matrices $U$ and $V^{\dagger}$ will also be real valued. In the case where the quantum state requires two physical qubits, both subsystems $A$ and $B$ will have one qubit each, so we would prefer if the operators $U$ and $V^{\dagger}$ could be directly implemented by ${\sf R}_y$ gates. This can be done directly if the determinant of the operator is $1$. If the determinant of the operator is $-1$, we can instead apply $V^{\dagger} \sf Z$ by an ${\sf R}_y$ gate and negate the angle of the ${\sf R}_y$ gate that prepares $\Sigma$ on subsystem $A$. The $\sf CNOT$ operation is converted to native neutral atom operations with the identity ${\sf CNOT} = ({\sf I}\otimes {\sf R}_y(\pi/2)){\sf CZ}({\sf I}\otimes  {\sf R}_y(-\pi/2))$. Thus, a circuit to prepare an arbitrary real two-qubit quantum state can be constructed using a single $\sf CZ$  gate and a few ${\sf R}_y$ gates as is shown in Fig. \ref{fig:2Q_Gray_Circuit}.

The circuit to prepare an arbitrary real three-qubit quantum state follows a similar construction (see Fig. \ref{fig:3Q_Gray_circuit}), except that sub-circuit $A$ has one qubit and sub-circuit $B$ has two. As a result, only the first two columns of $V^{\dagger}$ contribute to the final quantum state, so we elect to swap the third and fourth columns to make this matrix unimodular. An arbitrary unimodular two-qubit orthogonal gate can be implemented using at most two $\sf CZ$  gates, for a total $\sf CZ$ cost of three for this circuit \cite{Vatan2004}.

\begin{figure*}[!t]
    \centering
    \includegraphics[width=6.5in]{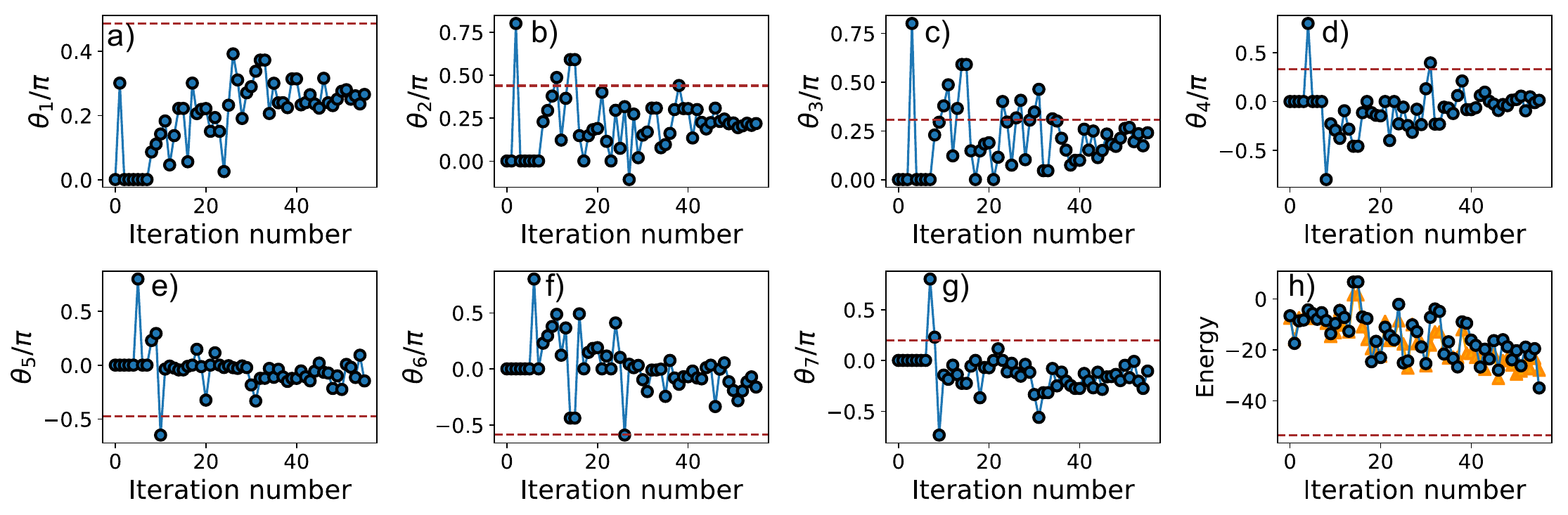}
    \caption{Results for $N=15$  spins encoded using 3 qubits with Gray code encoding. Panels a)-g) show the evolution of the variational parameters of the ansatz state during Nelder-Mead optimization. Panel h) shows the LMG energy during the optimization. The orange and teal points show the theoretical and experimentally measured energy for the parameter set used in each iteration. The horizontal dashed lines show the variational angles and energy corresponding to the ground state.}
    \label{fig:15spins}
\end{figure*}

\section{Simulation of 15 spin Hamiltonian }
\label{app.15spinVQE}
We attempted to simulate the ground state of $N=15$ spins encoded in 3 qubits with Gray code encoding. In order to simulate 15 spins, seven variational parameters are required, and the necessary basis transformations are the four Pauli string bases: $ \sf ZZZ,XZZ,ZXZ,ZZX$. The results are shown in Fig. \ref{fig:15spins}. We observed that the variational angles begin to converge, but except for $\theta_3$, all variational  angles (a-g) deviated from the theoretically calculated angles(shown with dashed  lines).  The energy in panel (h) also showed very weak convergence towards the ground state.

\rsub{The lack of convergence for the $N=15$ case is due to a number of factors including the dimensionality of the search space.} 
The $N=15$ circuit shown in Fig. \ref{fig:3Q_Gray_circuit} has the same width and gate depth as for the $N=9$ case but requires searching in a seven-dimensional space as opposed to the four-dimensional space for $N=9$. The convergence for $N=9$ was poor and required linear parameter scans and ZNE as seen in Figs. \ref{fig:9spins} and \ref{fig:9spinslinearscans} and Table \ref{tab:3spins}. \rsub{As demonstrated by the improvement in performance of the $N=9$ qubit case by ZNE, noise is another contributing factor. In the context of the VQE algorithm, there are two sources of noise. One is a consequence of imperfect quantum circuits and measurements. The second is due to the shot-noise inherent to the calculation of the expectation value for a given set of variational parameters. This shot-noise scales with the inverse square root of the number of bitstring measurements allocated to each circuit. We see a clear indication that circuit noise can cause a systematic difference between measured energy values and true values in the drift between the experimental and theory curves in later optimization iterations in Fig. \ref{fig:9spins}, which we expect is due to gate calibration drift. Gate calibration drift imposes constraints on the maximum iteration depth we can implement, as well as the number of bitstring measurements allocated to each set of variational parameters.} 

\begin{figure}[!t]
    \centering
    \includegraphics[width=\columnwidth]{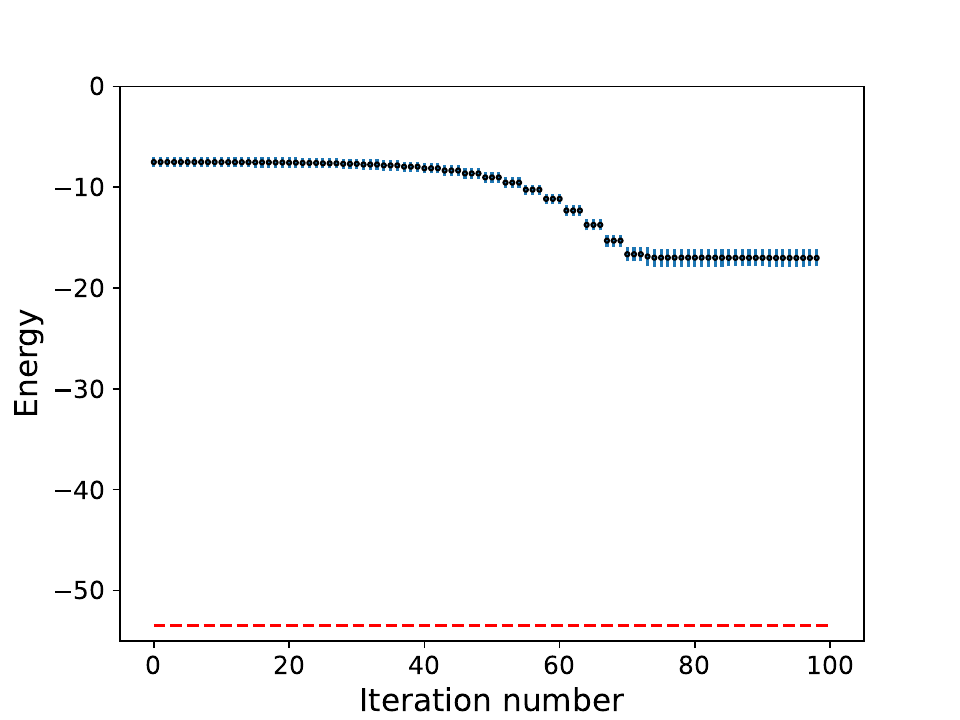}
    \caption{\rsub{Results for $N=15$  spins by direct classical evaluation of the energy of the ansatz state during Nelder-Mead optimization. The optimizer does not find the true ground state (dashed line) within the allocated number of iterations, even in the absence of circuit noise or measurement shot noise. }}
    \label{fig:noisefreesim}
\end{figure}

\rsub{ 
Figure \ref{fig:noisefreesim} shows the result of optimizing the $N=15$ case in the absence of both circuit noise and shot noise by directly evaluating the expectation value classically. Even in the absence of noise, the optimization did not converge to the optimal value of -53.47. The optimization was terminated at an iteration depth of 100, in excess of the iteration depths that each of the quantum optimizations were terminated due to run-time constraints imposed by gate calibration drift. To illustrate the consequence of shot noise at the level of 400 measurements per circuit used in the quantum simulations, we include error bars showing a one standard deviation range about the true mean. Even at one standard deviation, the shot noise would be large relative to the difference in energy values of adjacent iterations, which  dramatically limits an optimizer's ability to make progress.}

\rsub{In this noise-free optimization, each variational parameter was initialized to 0 to match the initial conditions of the quantum optimization. With an allowance for a greater maximum iteration depth, Nelder-Mead makes additional progress, but still fails to converge to the global optimum. This suggests that the outcome is sensitive to the initial variational parameters and the choice of the classical optimization algorithm. Collectively, these considerations imply that improved performance  would involve some combination of dynamically allocating bitstring measurements until enough strings have been gathered to reduce the standard error or the measurement below a desired threshold, running multiple optimizations with randomized initial variational parameters, and a potentially different choice of classical algorithm. These improvements come with increases in real-world runtime of the algorithm and their benefits must be considered accordingly.}

\section{Zero Noise Extrapolation (ZNE)}
\label{app.ZNE}
ZNE is an error mitigation technique widely used to find an expectation value in the presence of gate errors in the quantum processor. We take a generic noise-model-agnostic approach in which every noisy gate operator $\sf U$ is replaced by the operator $\sf U(U^{\dagger}U)$ \cite{Dumitrescu2018,He2020,Pascuzzi2022}. This is referred to in the main text as FIIM - ZNE. 
The additional inserted $\sf U^{\dagger}U$ operator is an identity operation for ideal gates, but for noisy operators, the experiential error increases each time this ``identity" is inserted into the circuit. In our experiment, the $\sf CZ$  gates are the most likely operations to cause an error, so we apply this strategy only to the $\sf CZ$  gates. One can add a quantity $n$ of identity insertions for each $\sf CZ$ gate, so each $\sf CZ$ gate will be replaced by $r=2n+1$ $\sf CZ$ gates. Assuming that the $\sf CZ$ gate error probability is small, the deviation in expectation value measured by the VQE algorithm from the exact noise-free value scales linearly with $r$. By measuring the cost function for different values of $r$, we can deduce the noise-free value occurring at $r=0$ using linear extrapolation. At $r=0$, we deduce what the energy would be if we could perform the circuit with zero $\sf CZ$ gates. In practice, performing the circuit without any $\sf CZ$ gates is impossible, so we cannot measure this value directly, but by linear extrapolation, we can estimate this value. We used FIIM - ZNE for 5 spin and 7 spin simulations (see Fig. \ref{fig:2Q_5S_7S} (c) and (f)).

If the circuit contains several $\sf CZ$ gates, it is possible for a single identity insertion to increase the error of the measured expectation values outside the linear regime. In this case, a different approach, SIIM - ZNE \cite{Pascuzzi2022}, can be used that applies identity insertions at only one $\sf CZ$ gate in the circuit at a time. The resulting noise-enhanced expectation value is then made for identity insertions at each of the $\sf CZ$ gates. The mean over the identity insertion locations is then used for linear extrapolation.  Since fewer $\sf CZ$ gate operations are added, this technique adds less noise for each identity insertion, allowing linear extrapolation to zero noise when the gate error is too large for FIIM - ZNE with the cost of requiring more expectation value measurements.  We used SIIM - ZNE for 9 spin simulations (see Fig. \ref{fig:9spinslinearscans}e).

\section{Experimental Apparatus}
\label{app.experiment}

The experimental platform is similar to that in Ref. \cite{Graham2022} with two major changes: the addition of top-hat beam shaping for the 459-nm single-qubit rotation and the Rydberg excitation beam and atom transport during quantum circuit execution. Initially, atoms were loaded into a blue-detuned line array trap \cite{Graham2019}, and a 1064-nm optical tweezer was used to ensure that the array sites required for the VQE calculation were loaded. The atoms were then transferred to a $2 \times 2$ 1064 nm Gaussian beam array created using crossed acousto-optic deflectors that were driven with two separate frequencies provided by a Quantum Machines OPX. Each trap in the array had a $1.2~ \mu$m beam waist and a depth of $600~ \mu$K. After recooling the atoms were optically pumped to initialize the qubit register; the final atom temperature was $15~ \mu$K. Each of the atoms was prepared in the correct state approximately $98.5\%$ of the time, and were measured with $99\%$ fidelity, taking these figures together yields a total state preparation and measurement (SPAM) error of $2.5\%$ per atom. 

During circuit execution, the atoms were reconfigured mid-circuit by ramping the frequencies with which the 1064-nm array's AODs were driven (see Appendix \ref{app.reconfigure}). This reconfiguration allowed gate operation provided by a combination of: 1. Microwaves for global $\sf{R}_\phi$ gates (where the rotation axis is located in the $x-y$ plane of the Bloch sphere at an angle $\phi$ from the $x$-axis). 2. a 459-nm beam (Vexlum VALO-SHG-SF) blue-detuned from the $7p _{1/2}$ level by 1.05 GHz to provide local $\sf Z$-rotations through the AC Stark shift. \footnote{\rsub{Note that the 1064-nm traps are left on during local $\sf Z$-gates. The traps shift both $6s_{1/2}$ and $7p_{1/2}$. However, the magnitude of this shift is small compared to the detuning of the 459-nm light from this transition, so the only way that such shifts affect the gates is to give a small change in gate duration. This affect is included in the gate calibration.}}  3. The same 459-nm beam and a 1040-nm beam (M-Squared SolsTiS-SA-PSX-XL) provide $\sf CZ$-gates. Together, these lasers could drive Rabi oscillations at up to $2\pi \times 6.5$ MHz; however, a lower Rabi frequency of $2\pi \times 1.7$ MHz was found to give better gate fidelity. We hypothesize that off-resonant Raman transitions out of the qubit basis were responsible for the lower fidelity at high Rabi frequencies. More details about the universal gate set can be found in the main text. 

\begin{figure*}[!t]
    \centering
    \includegraphics[width=.9\textwidth]{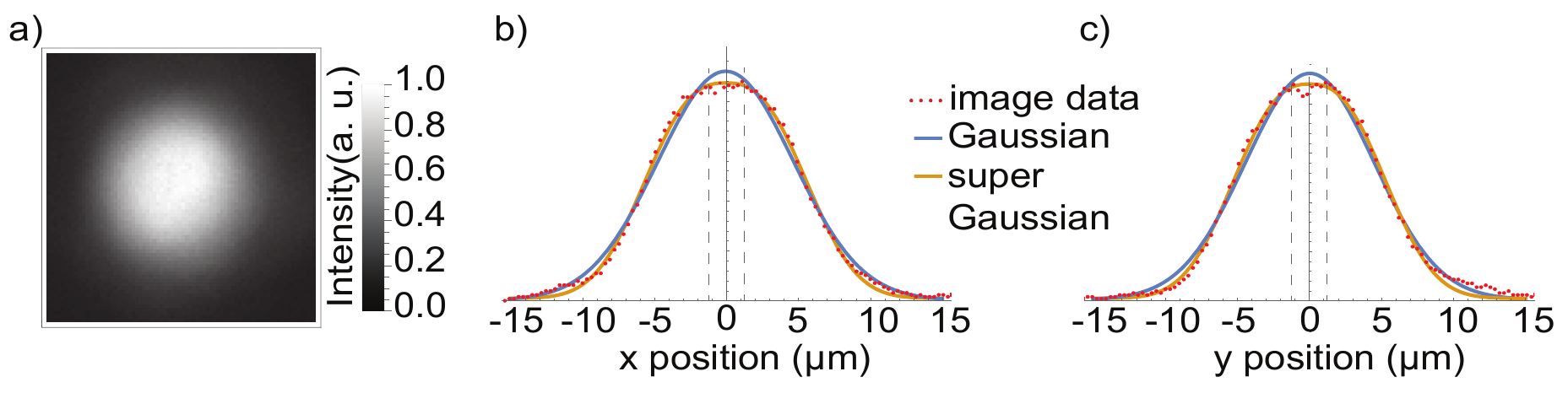}
    \caption{Characterization of the 459 nm beam.  a) The 2D beam intensity profile recorded on an inspection camera b) and c) show $x$ and $y$ cross sections of the beam fitted to Gaussian and super Gaussian ($p=2.6$) profiles with waists $w=8.8, 8.4~(\mu\rm m)$ for $x$ and $y$ cross sections respectively. The dashed lines show the atom positions with the beam centered symmetrically between them. }
    \label{SIfig:SiFig_Tophat}
\end{figure*}

A key factor in improving the fidelity of gate operations 2 and 3 was the top-hat beam shaping of the 459-nm light.
Because the Rydberg gates require both atoms to be simultaneously illuminated with the same beam, the Rydberg beams could not be simultaneously centered on both atoms when performing $\sf CZ$ gates. During rearrangement, the two targeted atoms are moved to a separation distance of $2.5~ \mu$m (both atoms located $1.25~ \mu$m from the center of the Rydberg beams). This displacement resulted in a non-zero intensity gradient on the atoms. This gradient and the uncertainty of the atom's position in the tweezer result in a shot-to-shot intensity fluctuation during the Rydberg gates. The Rydberg lasers drive the Rydberg excitation and shift the energy levels due to the AC Stark shift, so shot-to-shot intensity fluctuations result in quasi-static Rabi frequency and detuning fluctuations.  This noise can be reduced by increasing the Rydberg beam waist. Although this waist increase decreases the intensity gradient on the trapped atoms, it also decreases the intensity on the atoms and increases the amount of crosstalk to non-targeted sites. Despite the downsides, we use this technique for the 1040 nm Rydberg beam since we have an excess of power and crosstalk to neighboring sites does not significantly affect non-targeted sites as the 1040 nm light is far detuned from any transition out of the ground state. This allows us to use a Gaussian 1040 nm beam with a waist of $7.5~ \mu$m.

\begin{figure*}[!t]
    \centering
    \includegraphics[width=\textwidth]{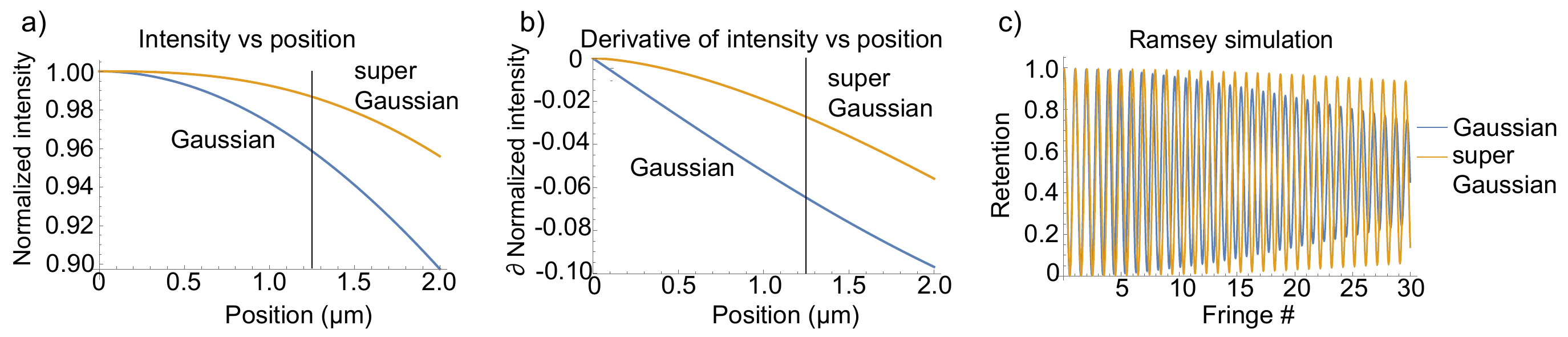}
    \vspace{-.5cm}
    \caption{Comparison of Gaussian and  super Gaussian beams with $p=2.6$ and waists $w=8.6~ \mu$m (see \ref{SIeq:SuperGauss2D}). a) and b) show the normalized intensity and its derivative of a Gaussian vs super Gaussian as a function of position; the super Gaussian beam has a more uniform intensity distribution at $1.25 ~\mu$m from the beam center where the atoms are trapped. c) Simulation of a Stark shifted Ramsey experiment showing reduced dephasing with a super-Gaussian beam. Parameters: 1064 nm trap with waist $w=1.2~\mu\rm m$, depth $k_B\times 600~\mu\rm K$, and atom temperature $15 ~\mu$K. The atom position was displaced 
    $d =1.25~ \mu$m from the beam centers. 
    }
    \label{SIfig:SiFig_GvsSG}
\end{figure*}

In contrast, the 459-nm Rydberg laser (though having an excess of power) does significantly affect non-targeted sites. Furthermore, it is more important for the 459-nm beam to have a small gradient on the atoms since it also is used for ${\sf R}_z$ gates where non-uniform intensity causes dephasing. We used beam-shaping to address these issues by placing a diffractive optical element (Holo/Or) to generate a top-hat beam profile upstream of the beam pointing AODs.   We imaged the 459-nm beam profile onto an inspection camera (see Fig. \ref{SIfig:SiFig_Tophat}). The resulting intensity distribution was well described by a super Gaussian of the form
\begin{align}
 I(x,y) &= I_0 e^{-\frac{2 \left[(x-x_0)^2+(y-y_0)^2 \right]^\frac{p}{2}}{ w^p}},
\label{SIeq:SuperGauss2D}
\end{align}
where $I_0$ is the maximum intensity of the super Gaussian beam, $(x_0, y_0)$ are the coordinates of the beam center, and $w$ is the beam waist. The parameter $p$ controls how flat the super Gaussian is; we were able to control this parameter in our imaging system by adjusting the beam size incident on the top-hat diffractive element. Since a super Gaussian beam has a flatter intensity profile, atoms displaced from the beam center see a smaller intensity variation and experience less dephasing due to Stark shift fluctuations (see Fig. \ref{SIfig:SiFig_GvsSG}). The $x$ and $y$ cross sections of the intensity distributions ($I(x,y_0)$ and $I(x_0,y)$) show a clear correspondence to a super Gaussian form.

We tested various values of $p$ by performing Ramsey experiments in the qubit basis with the traps in the $\sf CZ$ configuration. During this Ramsey experiment, we applied a 459-nm laser pulse. By varying the pulse time, we could measure the AC-Stark shift due to the 459-nm beam. We evaluated the intensity noise by measuring the number of coherent fringes in the oscillation ($f \tau$ where $f$ is the linear frequency of the oscillation and $\tau$ is the $1/e$ decay time). We found that $p=2.6$ gave the largest $f \tau$. Although only slightly flatter than a Gaussian beam ($p=2$), a super Gaussian with $p=2.6$ both theoretically and experimentally yielded an increase in $f \tau$ for a super Gaussian (83 theoretically and 30 experimentally) compared to a Gaussian (36 theoretically and 20 experimentally). We believe that the difference in the theoretical and experimental $f \tau$ numbers is due to imperfect beam quality.  A super Gaussian also reduces the crosstalk on the non-targeted site allowing neighboring atoms to be packed more densely without sacrificing performance. For a nearest-neighbor crosstalk threshold of $10^{-4}$, the atoms in a 2D array could theoretically be packed with a $43 \%$ higher density when using a super Gaussian with $p=2.6$ versus $p=2$ and the same beam waist. Theoretically, a super Gaussian with a larger $p$ should provide less intensity noise (with axial structure limiting $f\tau$ for a sufficiently large $p$) \cite{Gillen-Christandl2016}. However, we find that optical aberrations in the imaging line add axial structure to the beam focus for $p>2.6$ limiting the number of coherent fringes in the Ramsey experiment.

\section{Mid-circuit atom reconfiguration}
\label{app.reconfigure}

We have implemented reconfiguration during the circuit to enable low-crosstalk single-qubit gates when the atoms are far apart and strong interactions for $\sf CZ$ gates when the atoms are close together.   During the circuit, the atoms were trapped in a $2 \times 2$ array of red-detuned, Gaussian traps created with crossed AOD devices. The traps had waists of $w=1.2 ~\mu\rm m$, each with a depth of about $600 ~\mu\rm K$ giving a radial trap frequency of $\omega_0=2\pi\times 51~\rm kHz$. The sites were separated by $16 ~\mu\rm m$ for single-qubit operations and $2.5 ~\mu\rm m$ for $\sf CZ$ gates. 

The $d=13.5~\mu\rm m$ transport between near and far separations was implemented by frequency chirping of the AOD tones (generated by a Quantum Machines OPX)  to give a quintic transport profile with minimal jerk (derivative of acceleration) $x(t)=d\left[ 6(t/t_d)^5-15(t/t_d)^4+10(t/t_d)^3\right]$ for $0\le t \le t_d$ with a transport time $t_d=300~\mu\rm s$. The maximum acceleration was $a_{\rm max}=8.7\times 10^{-4}~\left(\mu\rm m/\mu\rm s^2\right)$.

We may compare the duration and motional heating of the minimum jerk profile with that of transport with constant jerk given by\cite{Bluvstein2022} 
$$
t_{\rm cj}=\frac{2^{1/4}3^{1/2}d^{1/2}}{\delta n^{1/4}x_{\rm ho}^{1/2}\omega_0}
$$
where $x_{\rm ho}=\left(\hbar/2m\omega_0\right)^{1/2}=  27~\rm nm$ for our trap parameters with Cs atoms, and $\delta n$ is the average increase of the vibrational quantum number. Following the same calculation\cite{Carruthers1965} for the minimal jerk profile we find
$$
t_{\rm mj}=\frac{2^{1/2}15^{1/3}d^{1/3}}{\delta n^{1/6}x_{\rm ho}^{1/3}\omega_0}.
$$
For a wide range of parameters, the minimal jerk profile provides less motional heating for the same transport time. 
Using the parameters given above, we find $\delta n_{\rm cj}=0.053$ and $\delta n_{\rm mj}=5.7\times 10^{-4}$, almost two orders of magnitude smaller.  With these parameters, less than $1 \mu$K of heating was observed after 5 round-trip ramps. The qubit coherence was preserved by applying \rsub{two} Hahn-echo pulses between round-trip transports to cancel accumulated phase shifts. \rsub{The first Hahn echo occurs after the Rydberg pulse, while the second takes place after the second transport. Each Hahn echo is a global microwave {\sf X} rotation, with each pulse lasting 8 $\mu$s. The entire sequence spans approximately 650 $\mu$s. In addition to canceling the accumulated phase shifts, the Hahn echo pulses mitigate the qubit dephasing during the transport sequence.}
 
\begin{figure*}
    \centering    \includegraphics[width=0.7\textwidth]{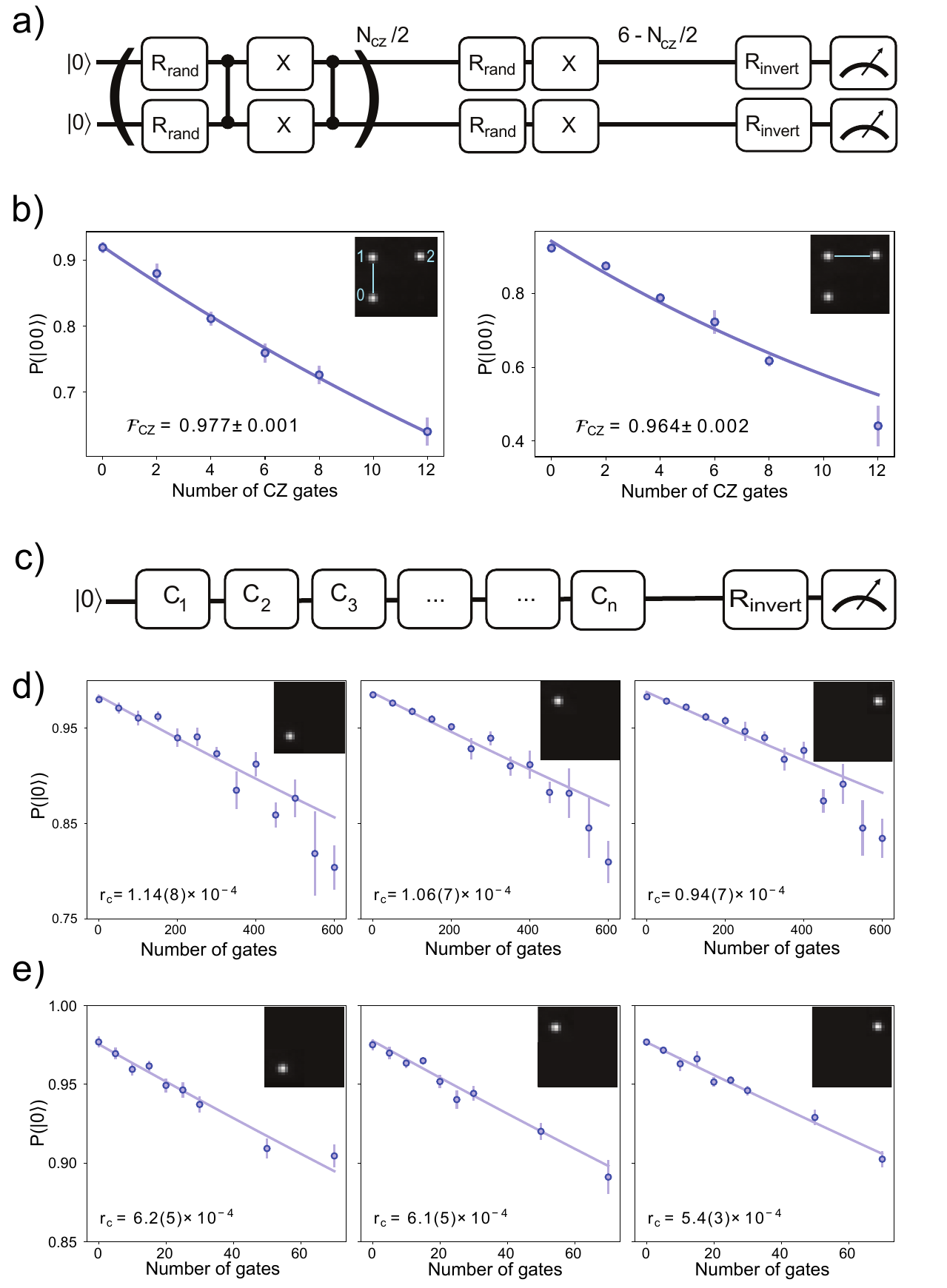}
    \caption{Quantum gate fidelity characterizations using Randomized benchmarking (RB).  a) Shows the RB circuit used to measure $\sf CZ$  gate fidelity. b) Shows the exponential decay of the two-atom ground state population as increasing number of $\sf CZ$ gates ($\sf CZ_{(q0,q1)}$ on the left and $\sf CZ_{(q1,q2)}$ on the right). The population at 0 $\sf CZ$ gates is limited primarily by SPAM and errors on the target two-atom state due to imperfect microwave rotations.  c) Shows the standard Clifford single qubit RB circuit used to characterize the global and local single-qubit gate sets. d) Shows the results of global microwave RB of all three sites. \rsub{$r_c=(1-\mathcal{F}_c)$ is the average error per Clifford gate, extracted from the decay fit \cite{Magesan2012b}.} e) Shows the results of RB of the local single qubit gates on all three sites.}
    \label{RB_results}
\end{figure*}

\section{Randomized benchmarking of quantum gates set}
\label{app.benchmarking}

We used randomized benchmarking (RB) to characterize the performance of the quantum gate set. 
 \rsub{After  applying a random gates sequence, followed by a final correction gate, the population in the $\ket{0}$ state  $P(\ket{0})_m$, is  fit to an exponential decaying function $P(\ket{0})_m=A_0 p^m+B_0$ where $m$ is the number of gates. The averaged error per Clifford gate $r_c$, is given by \cite{Magesan2012b}
\begin{equation}
    r_c=1-\mathcal{F}_c=\frac{(d-1)(1-p)}{d}
    \label{RB-fidelity-equation}
\end{equation}
where $d=2^n$ is the dimension of the $n$ qubit system.}

For the $\sf CZ$ gate fidelity characterization, we use the symmetric interleaved RB sequence demonstrated in ref. \cite{Evered2023}. Up to 12 $\sf CZ$  gates were interleaved with random single-qubit rotations applied globally with microwaves (see Fig. \ref{RB_results}a). The $\sf CZ$  gates are performed in pairs with an $\sf X$ gate inserted between to eliminate the single-qubit frame shift acquired during each $\sf CZ$  gate.
 We fit the data with an exponential decaying function to extract a gate fidelity of $\mathcal{F}_{{\sf CZ}(0,1)}=97.7(1)\%$ for qubit pair $q_0$ and $q_1$ and $\mathcal{F}_{{\sf CZ}(1,2)}=96.4(2)\%$ for qubit pair $q_1$ and $q_2$. \rsub{The fidelity difference observed between the two pairs is attributable to position-dependent intensity variation in the Rydberg beams. Each pair of atoms are illuminated by different portions of the beam. The $q_1-q_2$ atom pair experiences faster dephasing than $q_0-q_1$ due to a less flat intensity distribution when the atoms are horizontally spaced compared to when the atoms are vertically spaced.}

We used standard single-qubit Clifford randomized benchmarking \cite{Knill2008} to characterize both global and local single-qubit gates (see Fig. \ref{RB_results}c for a typical circuit). For global single-qubit rotations, we apply a global variant of the standard randomized benchmarking as shown in \cite{Xia2015} (see Fig.\ref{RB_results}d). \rsub{Using Eq. (\ref{RB-fidelity-equation}) with $d=2$}, we extracted an average gate fidelity of 0.99989(1) The faster decay after 500 quantum gates is due to changes in the qubit frequency due to a transient increase of the magnetic field.  A similar RB procedure was used to find the gate fidelity of a set of local Clifford gates constructed using global $\sf{R}_\phi$ and local $\sf{R}_z$ gates (see Fig.\ref{RB_results}e). We measured an average local Clifford gate fidelity of 0.99941(4).  

\section{Optimized $\sf CZ$ gate}
\label{app.bettergate}

In this section, we present the experimental modifications we made to improve the $\sf CZ$ gate fidelity \rsub{on the $q_1$ and $q_2$ atom pair}. These modifications were made after the VQE measurements were performed and highlight some strategies for fidelity improvement. These modifications include: 1. We excited to a lower Rydberg level, $55s_{1/2}$, which has a larger dipole matrix element compared to $66s_{1/2}$. The larger matrix element reduces the intensity required to drive the Rydberg transition at a given Rydberg Rabi rate. This effectively reduces infidelity caused by AC Stark shift fluctuations resulting from intensity fluctuations. 2. We found the optimal Rydberg Rabi frequency to be $2\pi \times 1$ MHz (versus $2\pi \times 1.7$ MHz in the VQE experiment). Although gates performed at lower Rabi frequency are more susceptible to some error sources such as Rydberg decay, the increase in such error sources is more than compensated for by the reduction in dephasing during the Rydberg excitation that was observed in the experiment. This dephasing reduction is not fully understood, but we believe that it is related to intensity noise found in the $459$-nm Rydberg laser. Additional measurements and simulations are needed to confirm this hypothesis. 3. After optical pumping, we added a trap power ramp to reduce the depth from $600 \mu$K to $260 ~\mu$K. This ramping, along with improved cooling parameters, reduced the atom temperature to $6 ~\mu$K. After these changes, we measured a $\sf CZ$ gate fidelity of $\mathcal{F}_{\sf CZ}=0.986(1)$ (see Fig.\ref{fig:RB_optimizeCZ}) using the methods described in Appendix \ref{app.benchmarking}.

\begin{figure}[!t]
    \centering    \includegraphics[width=0.5\textwidth]{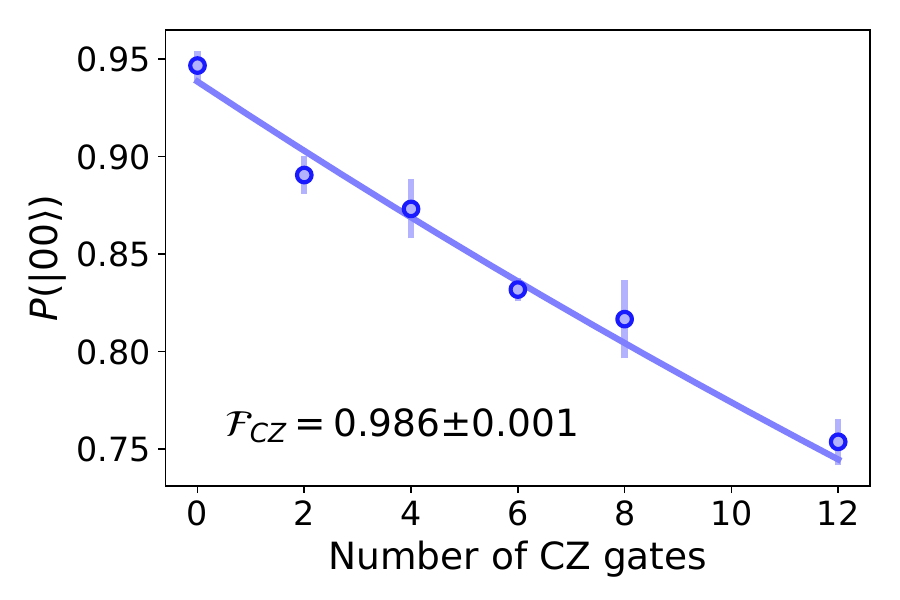}
    \caption{Result from randomized benchmarking of the optimized $\sf CZ$ gate giving fidelity of 98.6(1)\%. (see Appendix \ref{app.benchmarking} for more information)}
    \label{fig:RB_optimizeCZ}
\end{figure}

\bibliography{rydberg,optics,qc_refs,atomic,saffman_refs}

\end{document}